\documentclass[aps,prx,twocolumn,superscriptaddress,showpacs]{revtex4-1}

\usepackage{amsmath}
\usepackage{amsfonts}
\usepackage{amssymb}
\usepackage{graphicx}
\usepackage{color}
\usepackage{bm}
\usepackage{hyperref}
\usepackage{mathtools}
\usepackage{bbold}
\usepackage{empheq}
\usepackage[english]{babel}
\usepackage{pgfgantt}
\usepackage{dsfont}
\usepackage{enumerate}
\usepackage{comment}

\newcommand{\bra}[1]{\left\langle #1 \right\lvert}
\newcommand{\ket}[1]{\left\lvert #1 \right\rangle}

\renewcommand{\Im}{\mathop{\mathrm{Im}}}
\newcommand{\Tr}{\mathop{\mathrm{Tr}}}
\newcommand{\Ln}{\mathop{\mathrm{Ln}}}

\begin{document}

\title{Probing the topology of density matrices}
\author{Charles-Edouard Bardyn}
\affiliation{Department of Quantum Matter Physics, University of Geneva, 24 Quai Ernest-Ansermet, CH-1211 Geneva, Switzerland}
\affiliation{Kavli Institute for Theoretical Physics, UC Santa Barbara, USA}
\author{Lukas Wawer}
\affiliation{Department of Physics and Research Center OPTIMAS, University of Kaiserslautern, Germany}
\author{Alexander Altland}
\affiliation{Institut f\"ur Theoretische Physik, Universit\"at zu K\"oln, D-50937 Cologne, Germany}
\author{Michael Fleischhauer}
\affiliation{Department of Physics and Research Center OPTIMAS, University of Kaiserslautern, Germany}
\affiliation{Kavli Institute for Theoretical Physics, UC Santa Barbara, USA}
\author{Sebastian Diehl}
\affiliation{Institut f\"ur Theoretische Physik, Universit\"at zu K\"oln, D-50937 Cologne, Germany}
\affiliation{Kavli Institute for Theoretical Physics, UC Santa Barbara, USA}

\begin{abstract}
The mixedness of a quantum state is usually seen as an adversary to topological quantization of observables. For example, exact quantization of the charge transported in a so-called Thouless adiabatic pump is lifted at any finite temperature in symmetry-protected topological insulators. Here, we show that certain directly observable many-body correlators preserve the integrity of topological invariants for mixed Gaussian quantum states in one dimension. Our approach relies on the expectation value of the many-body momentum-translation operator, and leads to a physical observable --- the ``ensemble geometric phase'' (EGP) --- which represents a bona fide geometric phase for mixed quantum states, in the thermodynamic limit. In cyclic protocols, the EGP provides a topologically quantized observable which detects encircled spectral singularities (``purity-gap'' closing points) of density matrices. While we identify the many-body nature of the EGP as a key ingredient, we propose a conceptually simple, interferometric setup to directly measure the latter in experiments with mesoscopic ensembles of ultracold atoms.
\end{abstract}


\maketitle


\section{Introduction}
\label{sec:}

Topology has emerged as an important paradigm in the classification of ground states in many-particle quantum systems. Metaphorically speaking, topology enters the stage when the ground-state wavefunction of a complex quantum system contains ``twists'' (or ``knots''), as a function of defining system parameters. Relevant parameters sets may include, e.g., the  quasiparticle momenta labeling single-particle states in a translationally invariant system, the collective phase governing the macroscopic ground-state wavefunction of a superconductor, or the parameters controlling an external drive or pump.

Where topology is present, it is characterized by integer-valued invariants with a high degree of robustness with regard to perturbations, including parametric deformations of a Hamiltonian, translational symmetry breaking, or even the addition of particle interactions to symmetry-protected topological states in noninteracting systems. These invariants are generally formulated in terms of the zero-temperature \emph{ground state} of topological quantum systems, which implicitly assumes that the system can be described by a single pure-state wavefunction. Realistic systems, however, are generally characterized by mixed states corresponding to thermal, or even more exotic distributions. Therefore, an obvious set of questions presents itself: to what extent can the concept of topology be generalized to finite-temperature states and, more broadly, to arbitrary mixed states (described by a density matrix instead of a state vector)? And, even more ambitiously, are possible formal generalizations connected to topologically quantized observables?

The fact that density matrices lend themselves to topological classification is reflected in the definition of various geometric phases and corresponding topological invariants. An important example is given by the Uhlmann phase~\cite{Uhlmann}, a formal generalization of the geometric Berry phase~\cite{Berry, Simon, WilczekZee}. The definition of this phase is based on a gauge structure in the space of positive-definite density matrices~\cite{Uhlmann, Viyuela1D, Viyuela2D, Arovas2D, Budich2015}. Recently, a different approach has been proposed~\cite{Grusdt2017} to generalize the concept of topological order in the sense of Ref.~\cite{Chen2010} to mixed states, based on the equivalence of topologically identical states under local unitary transformations. While these approaches are formally elegant, they do not directly relate to observables that are readily representable in terms of system correlators~\footnote{Ref.~\cite{Viyuela2016} proposes to use an ancillary system in topological insulators that are simulated by single qubits to implement the formal state purification required to access the Uhlmann phase.}.

\begin{figure*}
    \includegraphics[width=\textwidth]{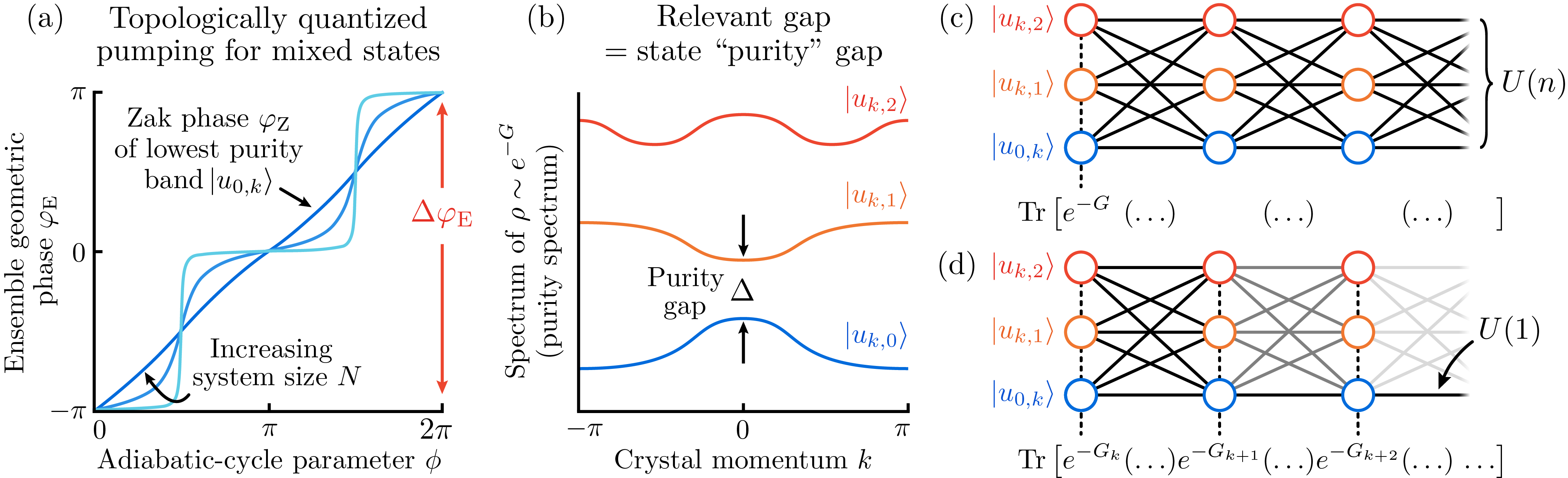}
    \caption{\textbf{Key results}. \textbf{(a)} Topologically quantized pumping for Gaussian mixed states with translation invariance. The ``ensemble geometric phase'' (EGP) $\varphi_\text{E}$ of a mixed state [Eq.~\eqref{eq:EGP}] reduces to the Zak phase of a pure state $\ket{u_{k, 0}}$ as the system size $N$ is increased. The state $\ket{u_{k, 0}}$ corresponds to the lowest band in the so-called ``purity spectrum'' [see (b)] of the state density matrix $\rho$. For thermal mixed states $\rho \sim e^{-\beta H}$, it coincides with the zero-temperature ground state of the Hamiltonian $H$. When varying a parameter $\phi$ along a loop from $0$ to $2\pi$, the phase $\varphi_\text{E}(\phi)$ changes by a topologically quantized value ($2\pi$ in the above illustration). This holds irrespective of $N$ and of the state mixedness. The only requirement is a gap in the state purity spectrum. \textbf{(b)} Schematic purity spectrum showing the relevant gap $\Delta$. The density matrix of the state can be expressed as $\rho = e^{-G}$, where $G = -\ln \rho$ is Hermitian [see, e.g., Eq.~\eqref{eq:GaussianDensityMatrix}]. This allows us to define the state ``purity'' eigenvalues and eigenvectors in analogy with spectral features of a Hamiltonian (where ``purity'' refers to the fact that eigenvalues indicate the degree of purity of $\rho$ in each eigenspace~\cite{Bardyn2013}). \textbf{(c), (d)} Crucial difference between conventional few-body observables and the many-body EGP considered in this work. While the former take the symbolic form $\sim \Tr[\rho(\ldots)] = \Tr[e^{-G}(\ldots)]$, the phase $\varphi_\text{E}$ has a more rigid structure $\sim \Tr[e^{-G_k}(\ldots)e^{-G_{k+1}}(\ldots) \, \ldots]$, where $G_k$ denotes the matrix $G$ in the momentum sector $k$. This structure corresponds to a path-ordered product over points $k = 1, \ldots, N$ across the Brillouin zone [Sec.~\ref{subsec:gaugeReductionMechanism}]. In the state eigenbasis $\{ \ket{u_{k, s}} \}$ (where $s = 0, 1, \ldots, n-1$ indexes purity bands), matrices $e^{-G}$ and $e^{-G_k}$ are diagonal and can be seen as ``weight factors''. In the conventional case depicted in (c) [where states $\ket{u_{k, s}}$ represented by circles are connected by operators $(\ldots)$], a single ``selection'' by weight factors occurs (indicated by a dashed line). For the EGP, in contrast, a thermodynamically large number of times $N$ of selections come into play. As a result, connections to states $\ket{u_{k, s}}$ with $s > 0$ are efficiently damped, and the EGP exhibits the $U(1)$ geometric properties of $\ket{u_{k, 0}}$ alone (up to corrections that vanish in the limit $N \to \infty$).}
    \label{fig:mainResults}
\end{figure*}

In this work, we explore a distinct notion of parallel transport for mixed states within the comparatively simple class of symmetry-protected topological (SPT) insulators. Our approach is conceptually different from previous works in that the starting point of our construction is a many-body correlator which is directly observable, instead of the entire density matrix. One of its defining features is its reduction to common classifying topologies in the zero-temperature limit where thermal mixed states become projectors onto quantum ground states. Our construction can therefore be regarded as an extension of such notions to finite temperatures or, more broadly, to ensembles of mixed states. Most importantly, it shows that topological quantization can survive mixedness at the cost of dealing with genuine many-body correlators. Specifically, focusing on translation-invariant one-dimensional (1D) lattice systems of fermions in Gaussian states, we identify a physical many-body observable $\varphi_\text{E}$ which approaches, in the thermodynamic limit, a well-defined geometric phase for mixed states. Although realistic distributions will often be generated by thermalization, the concept applies to generic density matrices (subject to a few conditions spelled out below). For this reason, we dub $\varphi_\text{E}$ the ``ensemble geometric phase'' (EGP).

Before describing our key results, we start by defining the aforementioned notion of topological ``twists'' in the context of SPT ground states. There, distinct topological sectors are identified as distinct homotopy classes characterizing how single-particle states $\{ | \psi_\alpha \rangle \}$ forming the many-particle ground state of a SPT insulator vary with parameter(s) $\alpha \equiv \{ \alpha_i \}$. Depending on the context, the relevant parameters may be ``internal'', such as the crystal-momentum components $\alpha_i \equiv k_i$ of a translationally invariant system, or ``external'', such as the parameters $\alpha_i \equiv \phi_i$ of an imposed drive or pump protocol. The homotopy classes characterizing the map $\alpha \mapsto | \psi_\alpha \rangle$ can be described in terms of a $U(n)$ Berry connection or gauge field $(A_i)_{ss'} \equiv i \langle \psi_{\alpha, s} | \partial_{\alpha_i} \psi_{\alpha, s} \rangle$, where $n$ is the number of bands (indexed by $s$) composing the ground state. In this picture, topological invariants can be understood as Chern classes of this gauge field (or quantities that depend on the latter~\cite{Ryu10}). More importantly, invariants are often related to (thus quantized) measurable observables, such as Hall transport coefficients. Where a direct connection to an observable exists, high levels of stability, e.g., with regard to disorder or particle interactions, are to be expected.

All these concepts are beautifully exemplified in the Rice-Mele model --- a paradigmatic model for noninteracting topological insulators in 1D~\cite{RiceMele} (with two bands and periodic boundary conditions; see Ref.~\cite{Asboth2016} for a recent review). In this model, a two-dimensional toroidal parameter space $\alpha \equiv (k, \phi)$ is defined by the system 1D Bloch momentum $k$ and an external pump parameter $\phi$. The relevant topological invariant then describes how many times the parameter-dependent ground state of the system completely covers an effective Bloch sphere during a full cycle of the parameters. When the external parameter $\phi$ is periodically and adiabatically varied in time, this invariant describes the quantized charge pumped through the system. Formally, the invariant is related to the Berry connection via a Chern class. It can be understood as the winding in $\phi$ space of the geometric (Berry) phase defined by the loop integral of the connection over $k$ (i.e., over the Brillouin zone), which in the 1D context is commonly called a Zak phase~\cite{Zak}. The Zak phase and the corresponding topological invariant are related to physical observables, i.e., to the zero-temperature (ground-state) polarization, and to the associated current flow.

\subsection*{Key results and outline}
\label{subsec:}

In this work, we consider the thermal equilibrium and nonequilibrium analogs of the afore-discussed noninteracting 1D systems, and extract topological information from their mixed states. The backbone of our construction is the expectation value of a many-particle momentum-translation operator~\cite{Linzner2016}, which was considered for pure states in a seminal work by Resta on the polarization of periodic systems~\cite{Resta1998}. Specifically, for reasons that we motivate in Sec.~\ref{sec:EGP}, we consider the ensemble geometric phase (EGP)
\begin{equation}
    \varphi_\text{E} \equiv \Im \ln \Tr \left( \rho e^{ i \delta k \hat{X}} \right),
    \label{eq:defegp}
\end{equation}
where $\rho$ is the relevant density matrix, $\delta k \equiv 2\pi/L$ is the smallest possible momentum in a periodic system of size $L$, and $\hat{X}$ is the many-particle (center-of-mass) position operator.

The EGP is a natural generalization of the geometric Zak phase relevant to pure quantum states. In particular, we demonstrate below that the value of the EGP for a mixed quantum state is given by the zero-temperature Zak phase of the ground state up to corrections that vanish in the thermodynamic limit. We thus find a positive answer to the question whether geometric and topological properties of pure quantum states may retain their integrity at finite temperatures or in general mixed quantum states. In fact, our construction identifies an \emph{order parameter for topological phase transitions in general mixed quantum states}. By studying concrete examples in and out of equilibrium, we demonstrate that topological phase transitions do exist in mixed quantum states, revealed by nontrivial (integer) changes in the winding of the EGP along a closed parameter cycle. We emphasize that single-particle quantities like the current, which can serve as topological order parameters at $T = 0$, are no longer related to a geometric phase at any finite temperature, and thus cannot provide information on the existence of topological phase transitions for mixed states. We will give the EGP topological order parameter a concrete physical meaning by specifying an experimental detection protocol based on many-body interferometry.

In more detail, in Sec.~\ref{sec:geometricPhase}, we investigate the geometric nature of the EGP in the context of noninteracting systems, focusing on mixed states generically described by a Gaussian density matrix $\rho \sim e^{-G}$. (We also consider translationally invariant lattice systems, for convenience.) All information about fermionic Gaussian states is encoded in the Hermitian matrix $G$ or, equivalently, in the covariance matrix $\Gamma$ collecting all expectation values of bilinears of fermion creation and annihilation operators. It will be convenient to think of $G$ as a ``fictitious Hamiltonian''. The spectrum of this Hamiltonian --- the ``purity spectrum'' --- describes the occupation probabilities of individual fermionic modes, and we will show that the existence of a gap in this spectrum --- a ``purity gap'' --- is required for the robustness of our construction [see Fig.~\ref{fig:mainResults}(b)].

In Sec.~\ref{subsec:gaugeReductionMechanism} and~\ref{subsec:twoBandExample}, we show that the EGP defines a geometric phase for mixed states in the thermodynamic limit, in the sense that
\begin{equation}
    \varphi_\text{E} = \varphi_\text{Z} + \Delta(N), \quad \Delta(N) \xrightarrow[]{N \to \infty} 0,
    \label{eq:keyResultEGPZakPhase}
\end{equation}
where $\varphi_\text{Z}$ is the Zak phase of the ``ground state'' --- or lowest ``purity band'' --- of the fictitious Hamiltonian $G$, and $\Delta(N)$ is a correction that vanishes in the limit of a large system size $N$ (number of unit cells) [see Fig.~\ref{fig:mainResults}(a)]. More explicitly, $\varphi_\text{Z}$ is defined as the loop integral over the Brillouin zone of the $U(1)$ gauge field or Berry connection of the lowest purity band $| u_{k, 0} \rangle$, i.e., $\varphi_\text{Z} = \oint dk A_k$, where $A_k \equiv i \langle u_{k, 0} | \partial_k u_{k, 0} \rangle$. Equation~\eqref{eq:keyResultEGPZakPhase} establishes the emergence, in the thermodynamic limit, of a bona fide $U(1)$ geometric phase for mixed quantum states. The underlying mechanism is discussed in Sec.~\ref{subsec:gaugeReductionMechanism} [and illustrated in Fig.~\ref{fig:mainResults}(c),(d)]. The EGP is defined up to integer multiples of $2\pi$ (i.e., it is a ``phase''), and its actual value on the unit circle is observable.

The above findings have an important consequence for EGP differences $\Delta \varphi_\text{E}$ accumulated over closed cycles in parameter space $(k, \phi)$ (where $\phi$ is an external parameter as above). Namely, we show that
\begin{equation}
    \Delta \varphi_\text{E} = \oint d\phi \, \partial_\phi \varphi_\text{E} = C,
\end{equation}
where $C$ is the Chern number associated with the lowest purity band $| u_{k, 0} \rangle \equiv | u_0 (k, \phi) \rangle$. This relation holds irrespective of the system size $N$, as the finite-size correction $\Delta(N)$ in Eq.~\eqref{eq:keyResultEGPZakPhase} cannot contribute to the integer winding of $\partial_\phi \varphi_\text{E}$. This demonstrates the existence of an exactly quantized observable for mixed quantum states, with an explicit connection to microscopic parameters of the system.

For pure states in the usual zero-temperature ground-state scenario, the correction $\Delta(N)$ in Eq.~\eqref{eq:keyResultEGPZakPhase} vanishes for any system size $N$, such that $\varphi_\text{E} = \varphi_\text{Z}$. Only in that case does the temporal variation $\partial_\phi \varphi_\text{E}$ of the EGP with respect to adiabatic changes of $\phi \equiv \phi(t)$ coincide with the physical charge current, and the difference $\Delta \varphi_\text{E}$ accumulated per adiabatic cycle correspond to the (quantized) transported charge. At finite temperature or in the nonequilibrium setting, in contrast, this connection breaks down. This can be understood from the fact that charge transport is related to the expectation value of a single-particle operator (the current), while $\Delta \varphi_\text{E}$ is related to a many-particle correlator [arbitrary powers of the single-particle operator $\hat{X}$ contribute to Eq.~\eqref{eq:defegp}]. This underlines the different phenomenology found in both cases: in charge transport, corrections to the current are intensive and finite in the thermodynamic limit, leading to the breakdown of exact quantization~\cite{Troyer2013, Nakajima2016}. In contrast, corrections to the EGP of thermal or nonequilibrium mixed states vanish in the thermodynamic limit, allowing for the strict quantization of $\Delta \varphi_\text{E}$ (to nontrivial values).

In Sec.~\ref{sec:examples}, we illustrate our general findings in minimal two-band models corresponding to the finite-temperature Rice-Mele model and to a nonequilibrium analog introduced in Ref.~\cite{Linzner2016}. In the thermal case, we demonstrate the existence of a topological phase transition (signaled by $\Delta \varphi_\text{E}$) at infinite temperature where the purity gap closes. In the other setting, we illustrate a chief feature of nonequilibrium dynamics, namely, the possibility that the purity gap closes at points where the ``damping gap'' of the Liouvillian describing the dynamics does not close~\cite{Bardyn2013}. Such singular points give rise to an observable nonzero $\Delta \varphi_\text{E}$ when encircled in parameter space, but are not associated with more conventional signatures such as divergent length and time scales in correlation functions.

Here, the requirement of adiabaticity in the conventional zero-temperature setting --- the smallness, as compared to some spectral gap (the Hamiltonian gap), of the rate of parameter changes along some path --- is replaced by a ``purity adiabaticity'' requirement: as detailed in Sec.~\ref{subsec:purityAdiabaticity}, the number of points at which the EGP must be measured or ``sampled'' along the relevant closed path in parameter space increases with decreasing purity gap. This reveals another analogy --- and important fundamental difference --- to the conventional zero-temperature setting: the adiabaticity condition comparing dynamical (energy or damping) scales is replaced, here, by a condition comparing dimensionless numbers.

The many-body nature of the correlator corresponding to the EGP [Eq.~\eqref{eq:defegp}] may shed doubts on its practical observability. In Sec.~\ref{sec:measurement}, however, we propose a scheme harnessing the tools available in current experiments to measure this phase in mesoscopic ensembles of ultracold atoms, via Mach-Zehnder interferometry using photons.

At this point, we would like to mention some more related works: Refs.~\cite{Avron11, Avron12} develop an adiabatic response theory for nonequilibrium (Liouvillian) dynamics for systems with few degrees of freedom, with recent generalization to a many-body context~\cite{Albert16}, where the connection to Hamiltonian ground-state responses is elucidated. This construction yields geometric phases and quantized invariants only in cases where the (instantaneous) stationary state is pure, or mixed in a specific fine-tuned way, in contrast to our situation. Furthermore, observable geometric phases have been identified in pumping protocols for open quantum dots in the high-temperature regime~\cite{Splettstoesser2017}. This construction builds up on ideas for geometric phases in classical dissipative systems~\cite{Landsberg1992}, which however do not relate to the geometric structure of the underlying quantum state.

\section{Resta polarization and its generalization}
\label{sec:EGP}

In this section, our goal is to construct the ensemble geometric phase $\varphi_\text{E}$, satisfying the following criteria:
\begin{enumerate}[(i)]
    \item The EGP is defined up to integer multiples of $2\pi$, i.e., it is a ``phase''.
    \item Differential changes $\partial_\phi \varphi_\text{E}(\phi)$ with respect to a parameter $\phi$ are well defined and observable.
    \item As a direct consequence of (i), the normalized change $\frac{1}{2\pi} \Delta \varphi_\text{E} \equiv \frac{1}{2\pi} \oint d\phi \, \partial_\phi \varphi_\text{E}(\phi)$ accumulated over a closed loop in parameter space is integer quantized.
    \item The EGP has a simple enough physical meaning for differences $\Delta \varphi_\text{E}$ to be measurable in a realistic setup.
    \item In limits where the relevant density matrix reduces to a projector onto a ground state, the EGP coincides with a conventional geometric phase (the Zak phase) characterizing ground-state band structures. A similar situation occurs when the EGP reduces to a projector onto an arbitrary pure state.
\end{enumerate}

In the following, we construct a quantity satisfying these conditions, and show how it defines a geometric phase for mixed states. We then identify the topological nature of its quantization property, and define a notion of quantized adiabatic pump for mixed states.

\subsection{Resta polarization}
\label{subsec:}

Key to our construction is a formulation of the electronic polarization of periodic quantum systems pioneered in an insightful paper by Resta~\cite{Resta1998}. Resta argues that the textbook expression $P = \langle \hat{X} \rangle / L = \langle \psi_0 | \hat{X} | \psi_0 \rangle$ (setting $e = \hbar = 1$) for the ground-state polarization of a 1D system with size $L$ in terms of the expectation value of the many-body position operator $\hat{X} \equiv \sum_j \hat{x}_j$ (where $\hat{x}_j$ is the position operator of individual particles $j$) is not applicable when the system is periodic as $\hat{X}$ is not a proper operator in the space of wavefunctions obeying periodic boundary conditions $\psi(L) = \psi(0)$. Instead, Resta suggests the alternative formula
\begin{equation}
    P = \frac{1}{2\pi} \Im \ln \langle \psi_0 | \hat{T} | \psi_0 \rangle, \quad \hat{T} \equiv e^{i \delta k \hat{X}},
    \label{eq:RestaPolarization}
\end{equation}
where $\delta k = 2\pi / L$. In this form, the polarization is defined modulo an integer (as required from the periodicity of the host system), and is expressed in terms of a \emph{phase}. Measurable incremental \emph{changes} $\Delta P$ are more relevant than the value of $P$ itself. In particular, the introduction of an adiabatically slow time-dependent parameter $\phi(t)$ leads to the observable $\partial_t P(t) = I(t)$ corresponding to the electronic current. Importantly, the charge $Q \equiv \Delta P = \oint dt \, I(t)$ ``pumped'' during a cyclic protocol $t \in [0, \tau]$ with $\phi(\tau) = \phi(0)$ is integer quantized. This fact follows in full generality from the observation that, for ground states $\ket{\psi_0(t)}$ that are weakly time dependent, the derivative $\partial_t P(t)$ is equal to an expression derived by Thouless and Niu~\cite{ThoulessNiu} for the current flowing in linear response to adiabatic changes. A more specific construction providing a connection to the topological band theory of noninteracting lattice systems considers $\ket{\psi_0}$ as a Slater determinant formed from single-particle Bloch states $\ket{\psi_{k, s}}$, where $k \in 2\pi \mathbb{Z}/N$ (for a system of length $L = Na$, with $N$ unit cells and lattice constant $a = 1$), and $s = 0, \ldots, n-1$ are band indices. Noticing that the many-body operator $\hat{T}$ introduced in Eq.~\eqref{eq:RestaPolarization} acts by translating all single-particle momenta by $\delta k$ (such that $k \to k - \delta k$), it is then straightforward to verify~\cite{Resta1998} that
\begin{align}
    P & = \frac{1}{2\pi} \Im \ln \prod_k \det(S_k) \nonumber \\
    & \simeq \frac{1}{2\pi} \oint dk \, \Tr(A_k) \equiv \frac{\varphi_\text{Z}}{2\pi},
    \label{eq:groundStatePolarization}
\end{align}
where $S_k$ is a $n_\text{p} \times n_\text{p}$ matrix (where $n_\text{p}$ is the number of particles in the system) formed by momentum-shifted ground-state wavefunction overlaps, namely, $(S_k)_{s, s'} \equiv \langle \psi_{k, s} | \hat{T} | \psi_{k, s'} \rangle = \langle \psi_{k, s} | \psi_{k - \delta k, s'} \rangle \simeq \delta_{s, s'} + i \delta k (A_k)_{s, s'}$, with $(A_k)_{s, s'} \equiv i \langle \psi_{k, s} | \partial_k \psi_{k, s'} \rangle$. The first expression in Eq.~\eqref{eq:groundStatePolarization} identifies $P$ as the (normalized) phase of a ``Wilson loop'' corresponding to the product of overlap determinants $\det(S_k)$ across the Brillouin zone. The second expression represents the same quantity as a discretized Berry phase, and the third provides a continuum approximation in terms of the Zak phase $\varphi_\text{Z} \equiv \oint dk \, \Tr(A_k)$ (Ref.~\cite{Zak}), which corresponds to the loop integral of the multiband $U(n)$ Berry connection $A_k$ (where $n$ is the number of bands).

The introduction of a time-dependent parameter $\phi \equiv \phi(t)$ leads to variations $\Delta P = \int dt \, \partial_t P = \oint d\phi \, \partial_\phi P$. When performing a full adiabatic cycle, one finds
\begin{align}
    \Delta P = \frac{i}{2\pi} \iint dk d\phi \Tr \left( \langle \partial_\phi \psi_0 |\partial_k \psi_0 \rangle - \langle \partial_k \psi_0 |\partial_\phi \psi_0 \rangle \right),
    \label{eq:pumpedCharge}
\end{align}
%
where the expression in the integral is a trace over the Berry curvature corresponding to the Berry connection $A_k$ (and its analog $A_\phi$ for variations in $\phi$). Equation~\eqref{eq:pumpedCharge} measures the integer homotopy invariant of the map $(\phi, k) \to | \psi_0 \rangle \equiv | \psi_0(k, \phi) \rangle$ from the torus defined by the two cyclic parameters $(\phi, k)$ to the ground-state manifold, i.e., $\Delta P$ corresponds to the number of times the ground-state wavefunction $| \psi_0(k, \phi) \rangle$ fully ``covers'' the torus in the process of a full parametric variation.

\subsection{Generalization to mixed states: the EGP}
\label{subsec:}

Taking advantage of the above formulation of the Zak phase of a pure state, we now turn to mixed states and construct a generalization of Eq.~\eqref{eq:RestaPolarization} (see also Ref.~\cite{Linzner2016}) designed to preserve all properties \mbox{(i)--(v)} above. We consider mixed states that arise as the unique stationary state of a gapped equilibrium or nonequilibrium quantum evolution --- the direct analog of gapped nondegenerate pure states~\cite{Bardyn2013}. The corresponding density matrix $\rho$ can be decomposed in the generic form $\rho = \sum_m p_m \ket{\psi_m} \bra{\psi_m}$, where $p_m > 0$ is the probability of finding the system in state $\ket{\psi_m}$. A natural generalization of the Zak phase $\varphi_\text{Z}$ [given by Eqs.~\eqref{eq:RestaPolarization} and~\eqref{eq:groundStatePolarization}] would be the average phase $\bar{\varphi}_\text{Z} = \sum_m p_m \varphi_{\text{Z}, m}$, where $\varphi_{\text{Z}, m}$ is the Zak phase of each individual pure state $\ket{\psi_m}$. This choice, however, trivially breaks property (i) above: the statistical average of phases defined modulo $2\pi$ is generally not defined modulo $2\pi$. Here, instead, we consider the phase of the statistical average $\sum_m p_m \langle \psi_m | \hat{T} | \psi_m \rangle$, i.e., we consider the ``ensemble geometric phase''
\begin{equation}
    \varphi_\text{E} =  \Im \ln \, \langle \hat{T} \rangle,
    \label{eq:EGP}
\end{equation}
where $\langle \ldots \rangle \equiv \Tr (\rho \ldots)$.
 
Equation~\eqref{eq:EGP} is designed to satisfy the benchmark criteria (i)--(v) above. In particular, we note that $\varphi_\text{E}$ reduces to a Zak phase $\varphi_\text{Z}$ in the limit of pure states, which hints at the geometric nature of the EGP for mixed states. Differential changes $\partial_\phi \varphi_\text{E}(\phi) \sim \Im (\langle \hat{T}\rangle^{-1} \partial_\phi \langle \hat{T} \rangle)$ are physically observable, as we will discuss in Sec.~\ref{sec:measurement}, and changes $\frac{1}{2\pi} \Delta \varphi_\text{E} \equiv \frac{1}{2\pi} \oint d\phi \, \partial_\phi \varphi_\text{E}(\phi)$ accumulated over parameter cycles are by construction integer quantized. For mixed states, $\frac{1}{2\pi} \Delta \varphi_\text{E}$ does not have the meaning of a pumped electric charge as in the case of pure states where it reduces to $\Delta P \equiv Q$ [Eqs.~\eqref{eq:groundStatePolarization} and~\eqref{eq:pumpedCharge}]. Nevertheless, we will show that it can be measured in many-body interferometric protocols.

In the following, we will demonstrate for a wide class of mixed states --- namely, \emph{Gaussian} mixed states --- that $\varphi_\text{E}$ is related to the Zak phase of a pure state up to corrections that vanish in the thermodynamic limit --- thereby establishing the topological nature of the quantization of $\frac{1}{2\pi} \Delta \varphi_\text{E}$. Gaussian mixed states can be represented by a quadratic Hermitian operator $\hat{G}$, and $\frac{1}{2\pi} \Delta \varphi_\text{E}$ corresponds, as we will show, to the ground-state topological invariant of the latter. In the specific case of thermal Gaussian states $\rho \propto e^{-\beta \hat{H}}$ --- the finite-temperature extensions of zero-temperature ground states --- the relevant operator is $\hat{G} = \beta \hat{H}$ (where $\beta$ is the inverse temperature), such that $\Delta \varphi_\text{E}$ reflects the topology of the ground state of $H$. Remarkably, for a gapped Hamiltonian $\hat{H}$, the quantity $\frac{1}{2\pi} \Delta \varphi_\text{E}$ remains quantized and coincides with the ground-state topological invariant $\Delta P$ of $\hat{H}$ at finite temperature, for as long as $\hat{G} = \beta \hat{H}$ is gapped (i.e., up to infinite temperature where $\beta = 0$). In this regard, the observable $\Delta \varphi_\text{E}$ reflects the zero-temperature ground-state invariant in a more robust way than single-particle observables (such as linear-response conductances) whose quantization is affected by intensive (system-size independent) ratios of band gaps over temperature (see, e.g., Refs.~\cite{Troyer2013, Nakajima2016}).

\section{Geometric phase and topological invariant for mixed states}
\label{sec:geometricPhase}

In this section, we explicitly relate the EGP defined in Eq.~\eqref{eq:EGP} to a geometric phase for mixed Gaussian states. To this end, we consider a set $\{ \hat{a}^\dagger_i, \hat{a}_i \}$ of fermionic creation and annihilation operators, where $i \equiv (r, s)$ is a composite index where $r = 0, \ldots, N - 1$ labels unit cells, and the band index $s = 0, \ldots, n - 1$ indexes fermionic sites in the unit cell (with lattice constant $a = 1$). We consider Gaussian states defined by a density operator of the form
\begin{equation}
    \rho = \frac{1}{\mathcal{Z}} \exp \left( -\sum_{i,j} \hat{a}^\dagger_i G_{ij} \hat{a}_j\right), 
    \label{eq:GaussianDensityMatrix}
\end{equation}
with $G$ is a Hermitian matrix and $\mathcal{Z}$ is a normalization constant ensuring that $\Tr(\rho) = 1$. The matrix $G$, which we call the state ``fictitious Hamiltonian'', plays a key role in what follows: it uniquely identifies the state, and its spectrum defines what we call the state ``purity spectrum''. The latter essentially corresponds to the spectrum of $-\ln \rho$, and its eigenvalues indicate the purity of the state in the corresponding eigenspaces~\cite{Bardyn2013}. Note that $G = \beta H$ for a thermal state $\rho \propto e^{-\beta \hat{H}}$ (where $H$ is the matrix representation of $\hat{H}$), as anticipated above.

The Gaussian density matrices that we focus on, defined in Eq.~\eqref{eq:GaussianDensityMatrix}, describe states that are fully characterized by single-particle correlations of the form $\langle \hat{a}^\dagger_i \hat{a}_j \rangle$, with vanishing ``anomalous'' correlations $\langle \hat{a}^\dagger_i \hat{a}^\dagger_j \rangle$ or $\langle \hat{a}_i \hat{a}_j \rangle$. Such states are typically found in equilibrium or nonequilibrium systems of noninteracting fermions without particle-number fluctuations. It is straightforward to check that operator expectation values calculated with respect to the density matrix in Eq.~\eqref{eq:GaussianDensityMatrix} are given by
\begin{equation}
    \langle \hat{a}^\dagger_i \hat{a}_j \rangle = [f(G)]_{ij}.
    \label{eq:correlationsFermiFunction}
\end{equation}
where $f(G) = (e^G + \mathbb{1})^{-1}$, with vanishing anomalous correlations. The same information is often collected in the so-called ``covariance matrix'' of the distribution, with matrix elements
\begin{equation}
    \langle [\hat{a}_j, \hat{a}^\dagger_i] \rangle = \left[ \tanh(\tfrac{G}{2}) \right]_{ij},
\end{equation}
Higher-order correlation functions can be calculated using Wick's theorem. Alternatively, and more efficiently here, one may compute the operator correlation functions of Gaussian states via a Grassmann integral, namely,
\begin{equation}
    \langle \hat{O}(\hat{a}^\dagger, \hat{a}) \rangle = \mathcal{N} \int d(\bar{\psi}, \psi) \, e^{\bar{\psi} f^{-1}(G) \psi} \hat{O}(\bar{\psi}, \psi),
\end{equation}
where $\hat{O}(\hat{a}^\dagger, \hat{a})$ is a (normal-ordered) operator defined in terms of the creation and annihilation operators, $\bar{\psi}_i$ and $\psi_i$ are Grassmann variables, $\hat{O}(\bar{\psi}, \psi)$ is obtained by the formal replacement $a_i^\dagger\to \bar \psi_i$, $a_i \to \psi_i$, and $\mathcal{N} = \mathrm{det}[-f(G)]$ normalizes the integral. We note that $\langle \hat{a}^\dagger_i \hat{a}_j \rangle \to \langle \bar{\psi}_i \psi_j \rangle = [f(G)]_{ij}$ readily follows from the Gaussian form of the integral, and the combinatorial signs in $\langle \hat{a}^\dagger_i \hat{a}^\dagger_k \hat{a}_j \hat{a}_l \rangle = -f(G)_{ij} f(G)_{kl} + f(G)_{il} f(G)_{kj}$ are faithfully reproduced by the Grassmann anticommutation $\psi_i \psi_j = -\psi_j \psi_i$.

We now use the above Grassmann representation to calculate the operator expectation value $\langle \hat{T} \rangle$ in Eq.~\eqref{eq:EGP}. To this end, we first express the operator in the form
\begin{align}
    \hat{T}(\hat{a}^\dagger, \hat{a}) & = e^{i \delta k \sum_i \hat{a}^\dagger_i x_i \hat{a}_i } \nonumber \\
    & = \prod_i \left(1 - \hat{a}^\dagger_i \hat{a}_i + \hat{a}^\dagger_i e^{i \delta k x_i} \hat{a}_i \right) \nonumber \\
    & = \prod_i e^{-\hat{a}^\dagger_i (1 - t_i) \hat{a}_i},
\end{align}
where we have defined $t_i = \exp(i \delta k x_i)$. The normal ordering of this expression is a straightforward operation as the indices $i$ carried by the factors entering the product $\prod_i$ are all different and, hence, nontrivial commutators do not appear. Next, we substitute $\hat{T}(\bar{\psi}, \psi) = \exp[-\sum_i\bar{\psi}_i (1 - t_i) \psi_i] \equiv 1 -\bar{\psi} (\mathbb{1} - T) \psi$, where $T \equiv \mathrm{diag}(t_i)$,
into the Grassmann integral to obtain
\begin{align}
    \langle \hat{T}(\hat{a}^\dagger, \hat{a}) \rangle \equiv \langle \hat{T} \rangle & = \mathcal{N} \int d(\bar{\psi}, \psi) \, e^{\bar{\psi} [f^{-1}(G) - \mathbb{1} + T] \psi} \nonumber \\
    & = \det[-f(G)] \det[-f^{-1}(G) + \mathbb{1} - T] \nonumber \\
    & = \det[\mathbb{1} - f(G) + f(G) T],
    \label{eq:TExpectFormal}
\end{align}
which is still a formal expression at this point, since $f(G)$ is a highly nondiagonal matrix in position space. A more tangible representation can be obtained by assuming that the state is translation invariant~\footnote{This requires both the dynamics and the initial state to be translation invariant~\cite{Rivas2013}.}, in which case the matrix $G$ can be prediagonalized in a ``Bloch basis'' as
\begin{equation}
    G = \sum_k G_k \ket{k} \bra{k}, 
\end{equation}
where $\langle x | k \rangle = N^{-1/2} e^{i x k}$. Here, $G_k$ is a $n \times n$ Hermitian matrix defined in band space, with elements $(G_k)_{s, s'}$, which we can be cast in diagonal form
\begin{equation}
    G_k = U_k B_k U_k^\dagger, \quad B_k \equiv \mathrm{diag}_s(\beta_{k, s}),
    \label{eq:stateMatrixDiagonalForm}
\end{equation}
where $U_k$ is a unitary matrix collecting the ``purity eigenstates'' (analogous to Bloch vectors) of the fictitious Hamiltonian $G_k$ representing the state, and $B_k$ is a diagonal matrix containing the ``purity eigenvalues'' $\beta_{k, s} \in \mathbb{R}$ of the latter, ordered in increasing order $\beta_{k, 1} \le \ldots \le \beta_{k, n}$, for convenience [see Fig.~\ref{fig:mainResults}]. The ``Fermi functions'' appearing in Eq.~\eqref{eq:TExpectFormal} can then be expressed as
\begin{equation}
    f(G) = \sum_k [U_k f(B_k) U_k^\dagger] \ket{k} \bra{k},
    \label{eq:FermiFunctionMomentumSpace}
\end{equation}
where $f(B_k) = (e^{B_k} + \mathbb{1})^{-1}$. The matrix $T$ in Eq.~\eqref{eq:TExpectFormal} (originating from the many-body translation operator $\hat{T}$) takes an intuitive form in Bloch basis: writing $x_i \equiv x_{r, s} \equiv x_r + x_s$, where $x_s$ is the position shift of the fermionic site $s$ located in the unit cell $r$ with position $x_r$, we obtain $\langle k | T | k' \rangle = \delta_{k, k' + 1}$~\footnote{\unexpanded{More precisely, one finds $\langle k | T | k' \rangle = \delta_{k, k' + 1} S$ with $S = \text{diag}_s(e^{i \delta k x_s})$. However, the matrix $S$ describing momentum shifts in the unit cell plays a very minor role here: it is diagonal in band space and can essentially be removed by a gauge transformation of the Bloch eigenstates of $G_k$, $\ket{u_{k, s}} \to (S^\dagger)^k \ket{u_{k, s}}$. This transformation leads to a modified boundary condition $\ket{u_{N, s}} = S^N \ket{u_{k, 0}}$, which is irrelevant for the geometric and topological properties of the EGP. We thus set $S = \mathbb{1}$, for simplicity.}}. [Note that we use $k$ interchangeably as a momentum index ($k = 0, \ldots, N - 1$) and as the momentum itself (with a factor $2\pi/N$).] Consequently, the matrix $T$ in Eq.~\eqref{eq:TExpectFormal} has the Bloch representation
\begin{equation}
    T = \sum_k | k + 1 \rangle \langle k |,
    \label{eq:TMatrixMomentumSpace}
\end{equation}
i.e., $T$ can be considered as a unit matrix on the diagonal next to the principal diagonal.

The EGP defined in Eq.~\eqref{eq:EGP} corresponds to the complex phase of $\langle \hat{T} \rangle$ in Eq.~\eqref{eq:TExpectFormal}. Since we are only interested in this phase, we can write
\begin{align}
    \varphi_\text{E} & = \Im \ln \det [\mathbb{1} - f(G) + f(G) T] \nonumber \\
    & = \Im \ln \det [\mathbb{1} + (\mathbb{1} - f(G))^{-1} f(G) T],
    \label{eq:EGPFormal}
\end{align}
where we have observed that $\ln \det [\mathbb{1} - f(G)] = \ln \det [\mathbb{1} - f(B)]$ is real and thus and does not contribute to the imaginary part of $\varphi_\text{E}$. Note that this assumes that the matrix $\mathbb{1} - f(G)$ is invertible, which is generically true for mixed states~\footnote{In the limit of pure states, one can always choose a proper regularization of $\mathbb{1} - f(G)$ to make it invertible.}. To reduce $\varphi_\text{E}$ to a more transparent form, we use Eqs.~\eqref{eq:FermiFunctionMomentumSpace} and~\eqref{eq:TMatrixMomentumSpace} along with the identity $\det(\ldots) = \exp \Tr \ln(\ldots)$ to write
\begin{align}
    \varphi_\text{E} & = \Im \ln \det [\mathbb{1} + (\mathbb{1} - f(B))^{-1} f(B) U^\dagger T U] \nonumber \\
    & = \Im \ln \left( \exp \Tr \ln [\mathbb{1} + (\mathbb{1} - f(B))^{-1} f(B) U^\dagger T U] \right) \nonumber \\
    & = \Im \ln \left[ \exp \Tr \ln (\mathbb{1} + M_T) \right] \nonumber \\
    & = \Im \ln \det (\mathbb{1} + M_T),
    \label{eq:phasePathOrderedProduct}
\end{align}
where the trace and determinant act in band space, in the last two lines, $U \equiv \{ \delta_{k, k'} U_k \}$, $B \equiv \{ \delta_{k, k'} B_k \}$, and $M_T$ is a path-ordered matrix product (``transfer matrix'')
\begin{align}
     M_T & \equiv (-1)^{N+1} \prod_k \frac{f(B_{k + 1})}{\mathbb{1} - f(B_{k + 1})} U^\dagger_{k + 1} U_{k} \nonumber \\
     & = (-1)^{N+1} \prod_k e^{-B_k} U_{k + 1, k},
     \label{eq:pathOrderedProduct}
 \end{align}
with ``link matrices'' defined as $U_{k + 1, k} = U^\dagger_{k + 1} U_k$. The matrices $U \equiv \{ \delta_{k, k'} U_k \}$ and $B \equiv \{ \delta_{k, k'} B_k \}$ appearing in Eq.~\eqref{eq:phasePathOrderedProduct} are block diagonal. In the crucial third equality in Eq.~\eqref{eq:phasePathOrderedProduct}, we have noted that $U^\dagger T U = \sum_k U^\dagger_{k + 1} U_k | k + 1 \rangle \langle k |$ is a matrix with blocks on the next-to-leading diagonal. Matrices of this form, viz. $A = \sum A_k | k + 1 \rangle \langle k |$, have the property that $A^N = (\prod_k A_k) \, \mathbb{1}$, while powers of $A$ different from multiples of $N$ do not contain diagonal matrix elements, and, hence, do not contribute to the expansion of expressions of the form $\Tr \ln (\mathbb{1} + A)$. This feature was used to arrive at the final expression in Eq.~\eqref{eq:phasePathOrderedProduct} containing the transfer matrix $M_T$. For convenience, we will assume an odd number of sites throughout, such that the factor $(-1)^{N+1}$ in Eq.~\eqref{eq:pathOrderedProduct} can be omitted.  

Equations~\eqref{eq:phasePathOrderedProduct} and~\eqref{eq:pathOrderedProduct} express the EGP as a path-ordered product of link matrices $U_{k+1, k}$ with points $k = 0, \ldots, N - 1$ along a closed loop corresponding to the Brillouin zone. This structure is reminiscent of discretized Wilson loop~\cite{Hatsugai2005}, where $U_{k+1, k}$ plays the role of a discrete $U(n)$ gauge connection. In the following, we will build on this observation to show how the EGP reduces to a more simple $U(1)$ gauge structure in the limit of large system sizes.

\subsection{Gauge-reduction mechanism}
\label{subsec:gaugeReductionMechanism}

Equations~\eqref{eq:phasePathOrderedProduct} and~\eqref{eq:pathOrderedProduct} directly relate the EGP to a path-ordered product of two types of matrices: the link matrices $U_{k + 1, k}$ and the ``weight factors'' $e^{-B_k} = \text{diag}_s(e^{-\beta_{k, s}})$. The link matrices describe the ``geometry'' underlying the band structure of the mixed state $\rho$. This can be understood from the fact that the $n \times n$ matrices $U_k \equiv (\ket{u_{k, 0}}, \ldots, \ket{u_{k, n}})$ contain the $k$-dependent Bloch eigenvectors diagonalizing the fictitious Hamiltonian $G_k$ representing the state [see Eq.~\eqref{eq:GaussianDensityMatrix}]. In the limit of large $N$ where points $k$ and $k + 1$ are infinitesimally close, the link matrices take the unitary form $(U_{k + 1, k})_{ss'} = \langle u_{k + 1, s} | u_{k, s'} \rangle \simeq 1 - \delta k \bra{u_{k, s}} \partial_k u_{k, s'}\rangle \simeq \exp[i \delta k (A_k)_{ss'}]$, where $A_k$ is the (non-Abelian) Berry connection
\begin{equation}
    (A_k)_{ss'} = i \bra{u_{k, s}} \partial_k u_{k, s'} \rangle.
\end{equation}
Therefore, the link matrices $U_{k + 1, k}$ describe the geometric ``twist'' of the state band structure (or ``purity'' bands) when moving from $k$ to $k + 1$ in the Brillouin zone.

The weight factors $e^{-B_k} = \text{diag}_s(e^{-\beta_{k, s}})$, on the other hand, are purely real and determine the statistical weight with which a given purity band $s$ contributes to the EGP. It is at this point that the many-body nature of the correlator $\langle \hat{T} \rangle$ in Eq.~\eqref{eq:EGP} really kicks in: while conventional few-body expectation values take the symbolic form $\sim \Tr[e^{-G}(\ldots)]$, i.e., a structure where weight factors appear once, here we have a much more rigid structure of the form $\Tr[\ldots (\ldots)e^{-G_{k + 1}}(\ldots)e^{-G_{k}} \ldots]$, in which selection through weight factors occurs a thermodynamically large number of times $N$ [see Fig.~\ref{fig:mainResults}]. For mixed states with a purity gap, weight factors efficiently select the lowest band $s = 0$ with weight $\beta_{k, 0}$ (for thermal states $\sim e^{-\beta \hat{H}}$, we recall that $\beta_{k, s} = \beta \epsilon_{k, s}$, where $\epsilon_{k, s}$ is the energy spectrum of the underlying Hamiltonian $\hat{H}$). In that case, Eq.~\eqref{eq:pathOrderedProduct} leads to a crucial gauge reduction to a single $U(1)$ component:
\begin{equation}
    \varphi_\text{E} \simeq \mathrm{Im} \ln \prod_k e^{-\beta_{k, 0}} e^{i \delta k A_{k, 0}}, \quad A_{k, 0} \equiv i \bra{u_{k, 0}} \partial_k u_{k, 0} \rangle.
    \label{eq:EGPApproximationGeometricPhase}
\end{equation}
As anticipated above, the relevant Berry connection for the EGP is now the $U(1)$ Berry connection $A_{k, 0}$ describing geometric properties of the lowest purity band.

As we will demonstrate below, corrections to Eq.~\eqref{eq:EGPApproximationGeometricPhase} generally vanish in the thermodynamic limit $N \to \infty$. In addition, differences $\frac{1}{2\pi} \Delta \varphi_\text{E} \equiv \frac{1}{2\pi} \oint d\phi \, \partial_\phi \varphi_\text{E}(\phi)$ accumulated per parameter cycle are topologically quantized irrespective of the system size $N$. The underlying gauge-reduction mechanism applies, in particular, to the thermal density matrices $\rho \sim \exp(-\beta \hat{H})$ of topological band insulators: while the probing of topological invariants via conventional response coefficients generally leads to compromised results for temperatures $\beta \Delta \epsilon \gtrsim 1$ exceeding the gap $\Delta \epsilon$ of $\hat{H}$, the observable $\frac{1}{2\pi} \Delta \varphi_\text{E}$ considered here remains topologically quantized at finite temperature. The only requirement is that the purity gap remains finite, i.e., $\beta \Delta \epsilon > 0$. In the next section, we will illustrate the reduction of the EGP to a geometric phase and the scaling of corrections in a simple example.

\subsection{Two-band example}
\label{subsec:twoBandExample}

In the following, we discuss the EGP in the illustrative case of a two-band model with gapped purity spectrum $\pm \beta_k \not = 0$. As a warmup, we first examine the case in which the density matrix reduces to a projector onto a pure state, which corresponds to the limit $\beta_k \equiv \beta \to \infty$. In this setting, the weight factors $e^{-(-\beta_k)} = e^{\beta} \to \infty$ of the lower purity band dominate over those of the upper band (with weights $e^{-\beta}$), and Eq.~\eqref{eq:EGPApproximationGeometricPhase} reduces to
\begin{align}
    \varphi_\text{E} & \simeq  \Im \ln \prod_k e^\beta e^{i \delta k A_{k, 0}} \simeq \Im \ln \left( e^{\beta N} \prod_k e^{i \delta k A_{k, 0}} \right) \nonumber \\
    & \xrightarrow[]{N \to \infty} \oint dk \, A_{k, 0} = \varphi_\text{Z},
    \label{eq:EGPReductionToZakPhasePureStates}
\end{align}
where all approximate equalities become exact in the limit $\beta \to \infty$. Note that the accumulated weight factors $\prod_k e^\beta = e^{\beta N}$ drop out as they do not contribute to the imaginary part. Therefore, in the pure-state limit, the EGP reduces to the Zak phase of the lower purity band, as expected~\footnote{Recall that, for thermal states $\rho \sim e^{-\beta H}$, the lower purity band corresponds to the ground state of $H$.}.

We now examine the case of mixed states, anticipating that Eq.~\eqref{eq:EGPReductionToZakPhasePureStates} will be reproduced up to corrections that vanish in the thermodynamic limit. The EGP is given by the path-ordered product defined by Eqs.~\eqref{eq:phasePathOrderedProduct} and~\eqref{eq:pathOrderedProduct}. We first parameterize the $2 \times 2$ unitary link matrices:
\begin{equation}
    U_{k + 1, k} \simeq e^{i \delta k A_k} \equiv e^{i \delta k \mathcal{A}_k \cdot \sigma},
    \label{eq:parametrizationLinkMatrices}
\end{equation}
where $\mathcal{A}_k \equiv (\mathcal{A}_k^0, \dots, \mathcal{A}_k^3)^T$ is a real four-component vector, and $\mathcal{A}_k \cdot \sigma \equiv \sum_{i = 0}^3 \mathcal{A}_k^i \sigma_i$ is a shorthand for the expansion of the $2 \times 2$ matrix $A_k$ in terms of Pauli matrices (with $\sigma_0 \equiv \mathbb{1}$). In the two-band setting examined here, the diagonal matrix $e^{-B_k} \equiv \text{diag}_s(e^{-\beta_{k, s}})$ in Eq~\eqref{eq:pathOrderedProduct} takes the form $e^{- \beta_k \sigma_3}$. For convenience, we decompose the link matrices in the form
\begin{align}
    U_{k+1, k} & = e^{i \delta k \sum_{i = 0, 3} \mathcal{A}_k^i \sigma_i} + \delta k \sum_{i = 1, 2} \mathcal{A}_k^i \sigma_i + \mathcal{O}(\delta k^2) \nonumber \\
    & \equiv U^\text{diag}_{k+1, k} + V_k,
    \label{eq:UvsA}
\end{align}
where $U^\text{diag}_{k+1, k} \equiv e^{i \delta k (\mathcal{A}_k^0 \sigma_0 + \mathcal{A}_k^3 \sigma_3)}$ is diagonal in band space, and $V_k \equiv \delta k (\mathcal{A}_k^1 \sigma_1 + \mathcal{A}_k^2 \sigma_2)$ causes transitions between bands. Defining $W_{k_1,k_2} = \prod_{k_2 \le k \le k_1} U^\text{diag}_{k+1, k} e^{-\beta_k \sigma_3}$, we may then expand the ``$\ln \det$'' in Eq.~\eqref{eq:phasePathOrderedProduct} perturbatively in transition matrix elements as
\begin{widetext}
    \begin{align}
        & \varphi_\text{E} = \Im \ln \det \left(G^{-1} + V^{(1)} + V^{(2)} + \ldots \right) = \Im \ln \det G^{-1} + \Im \Tr G V^{(2)} + \tfrac{1}{2} \Im \Tr (G V^{(1)})^2 + \ldots \, , \nonumber \\
        & \quad G^{-1} = \mathbb{1} + W_{N-1, 0} \, , \quad V^{(1)} = \sum_k W_{N-1, k} V_k W_{k-1, 0} \, , \quad V^{(2)} = \sum_{k > k'} W_{N-1, k} V_k W_{k-1, k'} V_{k'} W_{k'-1, 0} \, .
    \label{eq:EGPPerturbativeExpansion}
    \end{align}
\end{widetext}
One may read these expressions in the spirit of time-dependent perturbation theory, where $k$ plays the role of time, and $W_{k+1, k}$ corresponds to the unperturbed discrete time-dependent propagator. The presence of weight matrices in the ``Green's function'' $G$ implies that
\begin{align}
    G^{-1} & = \mathbb{1} + e^{i \delta k \sum_k \sum_{i = 0, 3} \mathcal{A}_k^i \sigma_i} e^{- \sum_k \beta_k \sigma_3} \nonumber \\
    & \approx \left( \begin{array}{cc}
        1 & 0 \\
        0 & e^{\bar{\beta} N} e^{i \delta k \sum_k A_{k, 0}}
    \end{array} \right),
    \label{eq:GreensFunctionExplicitForm}
\end{align}
where we have noticed that $\mathcal{A}_k^0 - \mathcal{A}_k^3 = i \langle u_{k,0} | \partial_k u_{k,0} \rangle \equiv A_{k, 0}$ is the $U(1)$ gauge field of the lowest purity band, and have defined the ``average purity'' of the lowest band $\bar{\beta} = N^{-1} \sum_k \beta_k$. The key point here is the exponential enhancement $\sim e^{\bar{\beta} N}$ with system size $N$ of the lower-band contribution relative to that of the upper band (with weight $\sim e^{-\bar{\beta} N}$). Inserting the above expression for $G^{-1}$ into Eq.~\eqref{eq:EGPPerturbativeExpansion} leads to a zeroth-order contribution which reproduces the result $\varphi_\text{E} = \varphi_\text{Z}$ obtained in Eq.~\eqref{eq:EGPReductionToZakPhasePureStates}. While the exponential factor  $e^{\bar{\beta} N}$ drops out upon taking the imaginary part of the logarithm, it does play a role in the experimental detection of the EGP. We will return to this point in Sec.~\ref{sec:measurement}.

Next, we examine perturbative corrections: in Eq.~\eqref{eq:EGPPerturbativeExpansion}, contributions with an odd number of interband-transition matrices $V_k$ vanish, and we focus on the leading second-order contribution. Referring to Appendix~\ref{sec:perturbativeCorrections} for details, we find that corrections to the zeroth-order result $\varphi_\text{E} = \varphi_\text{Z}$ scale as
\begin{equation}
    \Delta(N) = c \, [N \Delta \beta]^{-2},
    \label{eq:genericScalingCorrections}
\end{equation}
where $\Delta \beta = 2 \min_k{\beta_k}$ is the purity gap, and $c$ is a real constant independent of $N$ and $\Delta \beta$. Specifically, $\Delta(N) \sim (T/N)^2$ for thermal equilibrium states, implying that the EGP retains its zero-temperature value $\varphi_\text{E} = \varphi_\text{Z}$ even for finite $N$. Higher-order corrections vanish with higher powers of $N$ and, hence, are of negligible relevance. To understand the power-law suppression in an intuitive way, note that a thermodynamically nonvanishing (independent of $N$) correction can only arise if the smallness of the weight $\delta k \sim N^{-1}$ multiplying the $k$-local action of the transition operators $V_k$ gets compensated by an unconstrained summation over $k$ [since $\sum_k \delta k =\mathcal{O}(1)$]. However, excitations from $k$ to $k'$ of the (symbolic) form $\sum_{k, k'} (G_0)_{1 ,k} V_k (G_1)_{k, k'} V_{k'} G_{k', N} $ from the lower band ``$0$'' with ``propagator'' $G_0$ to the excited band ``$1$'' with propagator $G_1$ get weighed by a factor $\sim \exp(- \Delta \beta |k - k'|)$, on account of the spectral weights $e^{- \beta_k}$ of the density matrix penalizing excursions into the excited sector. This leads to ``confinement'' $|k - k'| \sim 1/(\Delta \beta)$, and implies that $\delta k^2 \sum_{k, k'} e^{-\Delta \beta |k - k'|} \sim 1/(\Delta \beta N)$. The fact that the correction actually scales as $1/(\Delta \beta N)^2$ with power two has to do with the fact that the leading-order perturbative expression comes out real, such that an additional factor $1/(\Delta \beta N)$ must be paid to obtain an imaginary contribution. The above mechanism applies regardless of the order of perturbation theory, and establishes the strong robustness of the geometric phase for mixed states defined by the EGP. Our perturbative calculations detailed in Appendix~\ref{sec:perturbativeCorrections} are supported by numerical simulations for various equilibrium and non-equilibrium models presented in Sec.~\ref{sec:examples}.

To summarize, we have shown that the EGP of purity-gapped fermionic Gaussian states with Bloch matrix representation $G_k$ satisfies $\varphi_\text{E} = \varphi_\text{Z} + \Delta(N)$, where $\varphi_\text{Z}$ is the Zak phase of the lowest purity band [given by Eq.~\eqref{eq:EGPReductionToZakPhasePureStates}], and $\Delta(N)$ is a correction that vanishes in the thermodynamic limit. For equilibrium thermal states $\rho \propto e^{-\hat{H}_k/T}$, the Zak phase is equal to $2\pi$ times the zero-temperature ground-state polarization of $\hat{H}_k$, and the observable $\varphi_\text{E}$ probes this value even at temperatures $T \sim \Delta \epsilon$ of the order or higher than the characteristic band gaps $\Delta \epsilon$ in the system.

We emphasize that the temperature dependence $\sim (T/N)^2$ of corrections in $\varphi_\text{E}$ is fundamentally different from that in single-particle observables probing topological quantization. In the latter case, corrections generally scale exponentially ($\sim e^{-\Delta \epsilon / T}$) with temperature $T \lesssim \Delta \epsilon$, independently of system size (see, e.g., Ref.~\cite{Rivas2013}), and crucially do not approach zero in the thermodynamic limit. The general mechanism identified here is also different from previous approaches focusing on Uhlmann-type phases~\cite{Viyuela1D, Viyuela2D, Arovas2D, Budich2015}: the latter are based on the construction of a system-size-independent geometric phase for density matrices, in contrast to the present construction where a gauge structure emerges only in the thermodynamic limit. Finally, we note that the scaling of corrections $\Delta(N)$ to $\varphi_\text{E} = \varphi_\text{Z}$ may be even more favorable in the presence of specific symmetries. An example is provided by the thermal density matrix of an SSH chain~\cite{SSH}: in that case, the sublattice (chiral) symmetry of the system leads to a Berry connection $\mathcal{A}_k$ in Eq.~\eqref{eq:parametrizationLinkMatrices} where one Pauli-matrix component is symmetry forbidden. Without going into detail, we mention that this symmetry leads to a correction $\Delta(N) \sim \exp(-\bar{\beta} N)$, exponential with system size.

\subsection{Topological nature of the quantized pumping}
\label{subsec:}

We have argued in Sec.~\ref{sec:EGP} that the EGP difference $\frac{1}{2\pi} \Delta \varphi_{\text{E}} = \oint
d\phi \, \partial_\phi \varphi_{\text{E}}(\phi)$ per cycle in some parameter $\phi$ is quantized in integer multiples of $2\pi$, which is not a priori obvious to reconcile with the finding that $\varphi_{\text{E}}$ is given by a geometric (Zak) phase plus a perturbative correction $\Delta(N)$. Since the EGP is defined modulo $2\pi$, however, the loop integral $\oint d\phi \, \partial_\phi \varphi_\text{E}(\phi) = \varphi_\text{E}(\phi_f) - \varphi_\text{E}(\phi_i)$ (where $\phi_i$ and $\phi_f \equiv \phi_i$ are the start and end points of the parameter cycle) must indeed be quantized in units of $2\pi$ irrespective of $N$. To reveal the topological nature and, hence, the robustness of this quantization, we first recall the above result that
\begin{equation}
    \varphi_\text{E}(\phi) = \Im \ln e^{i \oint_\text{BZ} dk A_0(k, \phi)} + \Delta(N),
\end{equation}
where $A_0(k, \phi)$ is the Berry connection of the lowest purity band (which here depends on $\phi$). Since $\Delta \varphi_\text{E}$ is quantized irrespective of $N$, the correction $\Delta(N)$ cannot contribute to its value. Instead, $\Delta \varphi_\text{E}$ is determined by the winding of the Zak phase $\varphi_\text{Z}(\phi) = \oint_\text{BZ} dk A_0(k, \phi)$ as $\phi$ is varied from $\phi_i$ and $\phi_f$. This winding formally corresponds to an integer topological invariant known as the Chern number, and we can write
\begin{eqnarray}
    \frac{1}{2\pi} \Delta \varphi_\text{E} = \frac{1}{2\pi} \iint d\phi \, dk \, F_0(k, \phi) = C \in \mathbb{Z},
    \label{eq:DeltaEGP}
\end{eqnarray}
where $F_0(k, \phi) \equiv \partial_\phi A_{k, 0}(k, \phi) - \partial_k A_{\phi, 0}(k, \phi)$~is the Berry curvature of the lowest purity band, defined in terms of the $U(1)$ gauge potential $A_{j, 0} \equiv -i \bra{u_0(k, \phi)} \partial_j \ket{u_0(k, \phi)}$ (where $\ket{u_0(k, \phi)}$ are the Bloch vectors forming the lowest purity band). Eq.~\eqref{eq:DeltaEGP} shows that $\Delta \varphi_\text{E}$ is a topologically quantized integer which coincides with the Chern number $C$ of the lowest purity band --- for any system size $N$ --- which is one of the key results of this work.

\subsection{Discussion}
\label{subsec:}

So far, we have illustrated our key gauge-reduction mechanism in a simple two-band model with a symmetric gapped purity spectrum ($\pm \beta_k \not = 0$ for all $k$). Here, we consider the more general case of $n$ bands and examine the role of the chemical potential (i.e., of band filling) in the thermal setting. We recall that the purity spectrum is given by $\beta_{k,s} = \beta \epsilon_{k,s}$ for thermal states $\sim e^{-\beta \hat{H}}$, where $\epsilon_{k,s}$ is the energy spectrum of $\hat{H}$ (and $s$ is the band index). If we work in the grand-canonical ensemble and introduce a chemical potential $\mu$, the relevant states become $\sim e^{-\beta (\hat{H} - \mu)}$, and the corresponding purity spectrum reads $\beta_{k,s} = \beta (\epsilon_{k,s} - \mu)$. Therefore, purity eigenvalues $\beta_{k,s}$ are positive (negative) for states located below (above) the chemical potential. In turn, the weight matrix $e^{-B_k} = \text{diag}_s(e^{-\beta_{k, s}})$ controlling the gauge-reduction mechanism in Eq.~\eqref{eq:pathOrderedProduct} contains exponentially decreasing (increasing) diagonal elements for states located below (above) the chemical potential. We can thus distinguish two cases depending on whether the chemical potential lies (i) within a gap, or (ii) inside a band (which would correspond to complete or partial filling, respectively, at $T = 0$):

In case (i), the direct analog of Eq.~\eqref{eq:EGPApproximationGeometricPhase} reads
\begin{equation}
    \varphi_\text{E} \simeq \mathrm{Im} \ln \prod_k {\prod_{s}}^\prime e^{-\beta_{k, s}} e^{i \delta k A_{k, s}}, \quad A_{k, s} \equiv i \bra{u_{k, s}} \partial_k u_{k, s} \rangle,
    \label{eq:EGPApproximationGeometricPhaseManyBands}
\end{equation}
where $\prod'_{s}$ denotes the product over bands $s$ located below the chemical potential (``filled'' bands). Accordingly, the EGP becomes $\varphi_\text{E} = \sum'_s \varphi_{\text{Z},s} + \Delta(N)$, where $\varphi_{\text{Z},s}$ is the Zak phase of the purity band $s$ and the sum runs over filled bands [as before, $\Delta(N)$ is a correction that vanishes in the thermodynamic limit]. The relevant topological invariant is then $\frac{1}{2\pi} \Delta \varphi_\text{E} = \sum'_s C_s$, where $C_s$ is the Chern number of the band $s$ located below the chemical potential. Therefore, in the general case of multiple bands, $\frac{1}{2\pi} \Delta \varphi_\text{E}$ reduces to its zero-temperature analog in a similar way as in the two-band model detailed above.

In case (ii) where the chemical potential lies within a specific band $s'$, the purity eigenvalues $\beta_{k, s'}$ change sign at certain values of $k$. As a result, the weight factors $e^{-\beta_{k, s}}$ only partially amplify the gauge-field contribution of the band $s'$, and $\varphi_\text{E}$ does not reduce to a sum of geometric (Zak) phases. In that case, as expected, $\frac{1}{2\pi} \Delta \varphi_\text{E}$ is not a topological invariant.

\subsection{Measurement of $\Delta \varphi_\text{E}$ and ``purity adiabaticity'' requirement}
\label{subsec:purityAdiabaticity}

In the conventional zero-temperature setting, where relevant states are pure (ground) states, topological order parameters can be determined by measuring currents integrated over a closed parameter cycle --- as typically envisioned in solid-state setups~\cite{ThoulessNiu} --- or, equivalently, by measuring the Zak-phase difference accumulated over a cycle --- as done in experiments with ultracold atoms~\cite{Bloch2013}. Such measurements rely on a \emph{dynamical} notion of adiabaticity, where pump parameters must be varied slowly in time as compared to the timescale set by some relevant gap (typically, the Hamiltonian gap).

Here we show that the requirements for measuring the mixed-state topological order parameter defined by the accumulated EGP difference $\frac{1}{2\pi} \Delta \varphi_\text{E}$ is more naturally related to a ``\emph{purity adiabaticity}'' criterion. To this aim, we propose to determine the topological invariant $\frac{1}{2\pi} \Delta \varphi_\text{E}$ from a set of $M$ independent measurements of EPG values $\varphi_\text{E}(\phi_j)$ at a discrete set of points $\phi_j$ along some relevant parameter cycle $\phi \in [0, 2\pi]$. The purity adiabaticity condition expressed in terms of the dimensionless purity gap $\Delta \beta$ and the dimensionless ``sampling rate'' (or inverse ``mesh size'') $\Delta \phi \equiv 1/M$ along the cycle (both assumed to be constant, for simplicity) then reads
\begin{equation}
    \Delta \phi \ll \Delta \beta,
    \label{eq:purityAdiabaticityCondition}
\end{equation}
to be contrasted to the usual dynamical adiabaticity criterion $\dot \phi \ll \Delta \epsilon$ relating the rate of parameter changes to an energy or damping gap (see also Appendix~\ref{sec:dynamicalAdiabaticity}).

To derive the above criterion, we examine how to extract the integer-quantized topological invariant $\frac{1}{2\pi} \Delta \varphi_\text{E}$ from a generically imperfect set of $M$ distinct EGP measurements $\varphi_\text{E}(\phi_j)$ along the relevant cycle in $\phi$ (where $j = 1, \ldots, M$). Following the approach of Ref.~\cite{Hatsugai2005}, we discretize the integral $\frac{1}{2\pi} \Delta \varphi_\text{E} = 1/(2\pi) \oint d\phi \, \partial_\phi \varphi_\text{E}$ in a way that crucially preserves two key properties: (i) the gauge invariance of $\frac{1}{2\pi} \Delta \varphi_\text{E}$, and (ii) its integer quantization. Specifically, we define the $U(1)$ ``link variables'' $U(\phi_j) \equiv \exp(i[\varphi_\text{E}(\phi_{j+1}) - \varphi_\text{E}(\phi_j)])$ and the corresponding ``lattice field strengths'' $F(\phi_j) \equiv \Ln U(\phi_j)$, where ``$\Ln$'' denotes the principal branch of the logarithm defined such that $-\pi < F(\phi_j)/i \leq \pi$. We then estimate the topological invariant of interest as the sum $\frac{1}{2\pi} \Delta \varphi_\text{E}' \equiv 1/(2\pi i) \sum_j F(\phi_j)$. Clearly, this quantity is invariant under gauge transformations $\varphi_\text{E}(\phi_j) \to \varphi_\text{E}(\phi_j) + 2\pi n_j$ (where $n_j$ is an arbitrary integer), and $\Delta \varphi_\text{E}' \to \Delta \varphi_\text{E}$ as $M \to \infty$. Remarkably, $\frac{1}{2\pi} \Delta \varphi_\text{E}'$ is additionally restricted, by construction, to integer values~\cite{Hatsugai2005}. As a result, one finds that $\Delta \varphi_\text{E}' = \Delta \varphi_\text{E}$ as long as the mesh size $M$ (or number of sampling points in parameter space) is larger than a critical size $M_c$. In fact, $\frac{1}{2\pi} \Delta \varphi_\text{E}'$ can only change (i.e., jump by an integer value) when $|F(\phi_j)| = \pi$ at some point $j$ in parameter space, which corresponds to a large discontinuity $|\varphi_\text{E}(\phi_{j+1}) - \varphi_\text{E}(\phi_j)| = \pi$ [modulo $2\pi$, as $2\pi$ jumps do not contribute to $U(\phi_j)]$. Accordingly, the critical mesh size can be estimated as the size below which the ``admissibility condition'' $|F(\phi_j)| < \pi$ (for all $j$) breaks down.

In summary, the value $\frac{1}{2\pi} \Delta \varphi_\text{E}'$ extracted from independent EGP measurements via the above procedure \emph{exactly} coincides with the integer topological invariant $\frac{1}{2\pi} \Delta \varphi_\text{E}$ \emph{provided} that $\varphi_\text{E}(\phi)$ is measured at a sufficiently large number of points $M > M_c$. In general, the critical mesh size $M_c$ is controlled by the proximity of the cyclic path $\phi \in [0, 2\pi]$ to gap closing points (see, e.g., Ref.~\cite{Dauphin2016}): the latter can be seen as sources of Berry-type curvature, in the sense that the field strength $F(\phi)$ is concentrated at such points in the limit of an infinitesimal mesh $M \to \infty$. Here, the relevant gap is the purity gap. Indeed, as we have demonstrated in Sec.~\ref{sec:examples}, the EGP winding $\frac{1}{2\pi} \Delta \varphi_\text{E}$ vanishes for parameter cycles that do not encircle one (or more) purity-gap-closing point(s). This allows us to define the above notion of ``purity adiabaticity'' unique to thermal and nonequilibrium systems: to be able to observe the topological invariant $\frac{1}{2\pi} \Delta \varphi_\text{E}$, one must sample a number $M > M_c$ of EGP values which gets larger and larger as one approaches purity-gap-closing points --- diverging exactly at such points~\footnote{The critical mesh size $M_c$ increases with the absolute value of the topological invariant $\frac{1}{2\pi} \Delta \varphi_\text{E}$ [i.e., with the amplitude of the curvature concentrated at the gap-closing point(s)]. The critical value $M_c$ is typically not large in systems where $\Delta \varphi_\text{E} \sim \mathcal{O}(1)$~\cite{Dauphin2016}.}. This leads to the criterion presented in Eq.~\eqref{eq:purityAdiabaticityCondition}.

We emphasize that measurement errors on the discrete values $\varphi_\text{E}(\phi_j)$ are irrelevant as long as the admissibility condition $|F(\phi_j)| < \pi$ (for all $j$) remains satisfied (recall that $\Delta \varphi_\text{E}'$ cannot change without breaking this condition). Therefore, errors can generically be compensated for by (i) using a finer mesh, or/and (ii) choosing parameter cycles further away from purity-gap-closing point(s).

Unlike usual measurements of accumulated Zak phase differences in the zero-temperature setting, the above procedure for measuring $\Delta \varphi_\text{E}$ does not rely on any dynamical protocol. This provides intuition as to why the purity adiabaticity criterion in Eq.~\eqref{eq:purityAdiabaticityCondition} involves a comparison of dimensionless numbers instead of dynamical scales. Since the values $\varphi_\text{E}(\phi_j)$ can be determined via completely independent measurements, the system can always be prepared with fixed parameters and measured after the time required for reaching its stationary state (controlled by possibly complex thermalization processes, in the thermal Hamiltonian case, or by a given damping gap, in the nonequilibrium Liouvillian case) --- with otherwise no requirement for adiabaticity under dynamical changes of parameters. For completeness, however, we present in Appendix~\ref{sec:dynamicalAdiabaticity} a detailed analysis of dynamical quasi-adiabatic measurements of $\Delta \varphi_\text{E}$, where parameters are varied continuously in time. The advantage of such measurements as opposed to independent ones as above is that the state of the system follows the quasi-adiabatic evolution of parameters, which naturally fixes the gauge and leads to continuous changes in $\varphi_\text{E}(\phi)$. At the end of the parameter cycle, the topological invariant $\frac{1}{2\pi} \Delta \varphi_\text{E}$ is simply given by $\frac{1}{2\pi} |\varphi_\text{E}(\phi = 2\pi) - \varphi_\text{E}(\phi = 0)|$. The downside, however, is that this value generically deviates from an integer, due to the dynamical errors that come into play (see Appendix~\ref{sec:dynamicalAdiabaticity}).

\section{Equilibrium (thermal) and non-equilibrium examples}
\label{sec:examples}

In this section, we demonstrate our analytical results numerically in two illustrative examples: (i) the Rice-Mele model in thermal equilibrium, and (ii) its nonequilibrium driven-dissipative analog introduced in Ref.~\cite{Linzner2016}. Both models are noninteracting and translationally invariant. They exhibit Gaussian states $\rho \sim e^{-G}$ [Eq.~\eqref{eq:GaussianDensityMatrix}], described by a ``fictitious Hamiltonian'' $G$ (or $G_k$, in momentum space), and the EGP can be computed, e.g., using the path-ordered formula found in Eq.~\eqref{eq:phasePathOrderedProduct} and~\eqref{eq:pathOrderedProduct}. We will illustrate three key features: (i) the convergence, in limit of large system sizes, of the EGP $\varphi_\text{E}$ to the Zak phase $\varphi_{\text{Z}}$ of the lower band of $G_k$ (the lower purity band), (ii) the quantization of the EGP difference $\Delta \varphi_\text{E}$ accumulated over a closed cycle in parameter space, and the coincidence of $\tfrac{1}{2\pi} \Delta \varphi_\text{E}$ with the Chern number of the lower purity band, and (iii) the direct connection between purity-gap-closing and topological transitions in $\tfrac{1}{2\pi} \Delta \varphi_\text{E}$.

We first examine the Rice-Mele model~\cite{RiceMele}, defined by the Hamiltonian
\begin{align} 
    \hat{H} = & \sum_r \left( t_1 \hat{a}_{r, 1}^\dagger \hat{a}_{r, 0} + t_2 \hat{a}_{r + 1, 0}^\dagger \hat{a}_{r, 1} + \mathrm{H.c.} \right) \nonumber \\
    & - \Delta \sum_{r, s} (-1)^s \hat{a}_{r, s}^\dagger \hat{a}_{r, s},
    \label{eq:RiceMeleModel}
\end{align}
where $r = 0, \ldots, N - 1$ indexes unit cells and $s = 0, 1$ indexes fermionic sites in the unit cell. The first line describes the hopping of fermions on a 1D lattice with alternating hopping amplitudes $t_1$ and $t_2$, and the second line describes a staggered potential. At $\Delta = 0$, the model reduces to the SSH model~\cite{SSH}. It exhibits chiral symmetry, which promotes the Zak phase to a topological invariant. In that case, two topologically distinct phases (protected by chiral symmetry) can be distinguished for $t_1 > t_2$ and $t_1 < t_2$, respectively (separated by a gapless point at $t_1 = t_2$). The corresponding quantized values of the Zak phase are $\varphi_\text{Z} = 0$ and $\pi$ (modulo $2\pi$), respectively, which corresponds to ground-state polarizations $P = \varphi_\text{Z}/(2\pi) = 0$ and $1/2$. In the Rice-Mele model, the parameter $\Delta$ provides a way to break chiral symmetry and, hence, to adiabatically connect the two phases originating from the SSH model and induce quantized polarization changes $\Delta P$ (note that the two phases are not symmetry protected anymore when $\Delta \neq 0$). In particular, adiabatic cycles in parameter space $(t_1 - t_2, \Delta)$ lead to an integer-quantized polarization difference (i.e., a pumped charge) $\Delta P = 1$ whenever the gapless point $t_1 = t_2$, $\Delta = 0$ is encircled. This process corresponds to a topological (Thouless) pump.

We now examine the behavior of the Rice-Mele model at finite temperature where the EGP replaces the Zak phase as the relevant probe for topology. The fictitious Hamiltonian representing the thermal state $\rho \sim e^{- \beta H_k} \equiv e^{- G_k}$ of the system is given by $G_k = \beta H_k$, where $H_k$ is the momentum-space Hamiltonian matrix $H$ corresponding to Eq.~\eqref{eq:RiceMeleModel}. It can be expressed in the form
\begin{align}
    \begin{split}
    G_k & = \, \mathbf{n}_k \cdot \boldsymbol{\sigma} \equiv \beta_k U_k \sigma_3 U_k^\dagger, \\
    \mathbf{n}_k & = \beta (t_1 + t_2 \cos k, t_2 \sin k, -\Delta)^T, \,\, \beta_k = || \mathbf{n}_k ||,
    \label{Eq:RiceMele1stQuantized}
    \end{split}
\end{align}
where $\boldsymbol{\sigma} = (\sigma_1, \sigma_2, \sigma_3)^T$ is a vector of Pauli matrices. The matrix $G_k$ is diagonalized by the unitary matrices $U_k$, and its spectrum (the purity spectrum) takes the form $\pm \beta_k$ with $\beta_k = \beta \epsilon_k$, where $\epsilon_k$ is the energy spectrum of the underlying Hamiltonian $H_k$ [using similar notations as before, as in Eq.~\eqref{eq:stateMatrixDiagonalForm}].

\begin{figure}
    \includegraphics[width=\columnwidth]{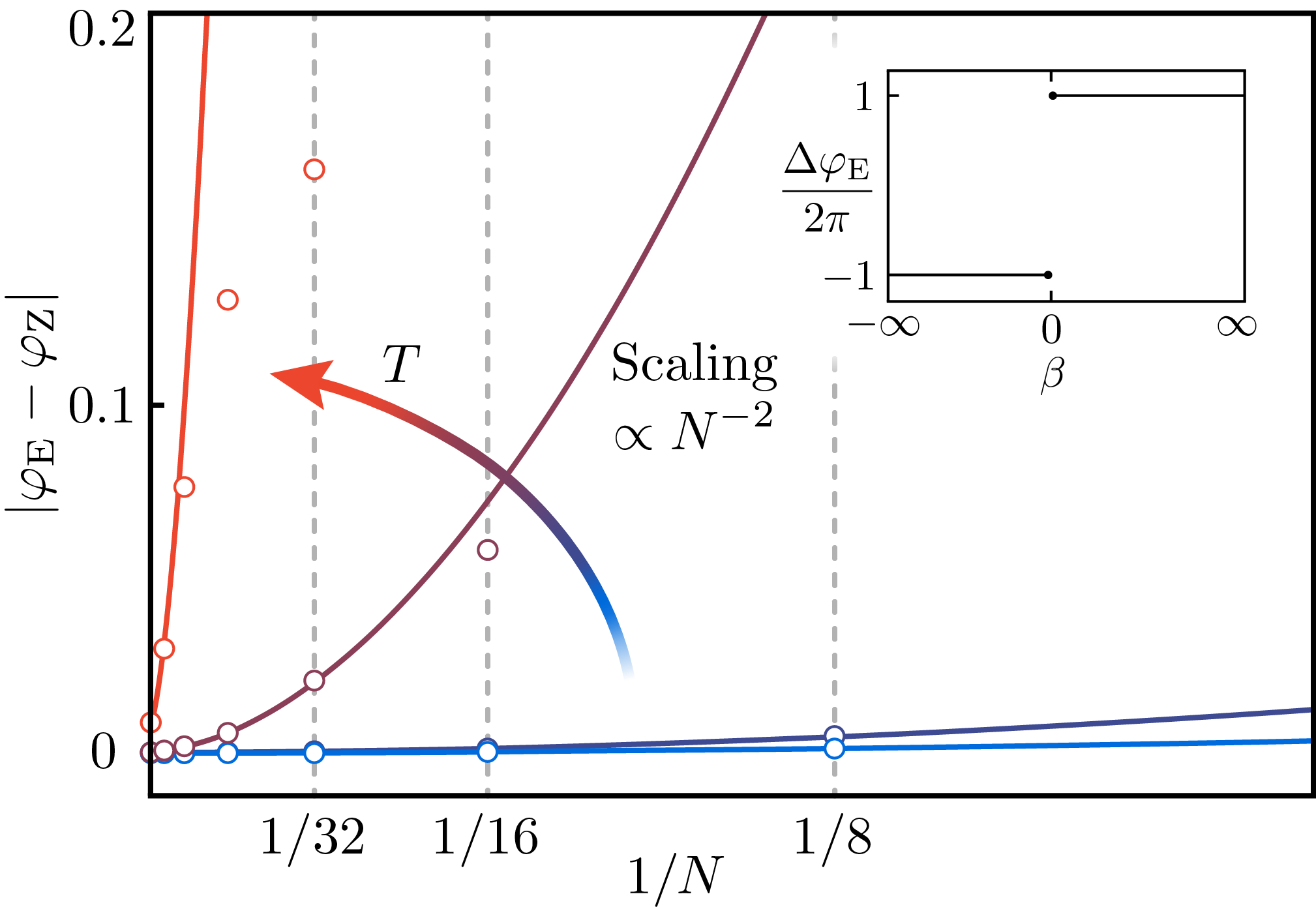}
    \caption{Scaling of finite-size corrections $| \varphi_\text{E} - \varphi_\text{Z} |$ in the finite-temperature Rice-Mele model, for different temperatures $T = 0.1, 1, 10,$ and $100$, with $t_1 - t_2 = -1/4$ and $\Delta = \sqrt{3}/4$ in Eq.~\eqref{eq:RiceMeleModel}. Solid lines are fitting curves $\sim N^{-2}$, which provide a good description of data points in the limit of large system sizes $N$. \textbf{Inset}: Difference $\tfrac{1}{2\pi} \Delta \varphi_\text{E}$ accumulated per closed cycle in $(t_1 - t_2, \Delta)$ space encircling the purity-gap-closing point $t_2 = t_1$, $\Delta = 0$, as a function of inverse temperature $\beta$. The same plot is found for $N = 8, 16, 32$, showing that the quantization of $\tfrac{1}{2\pi} \Delta \varphi_\text{E}$ is independent of system size. As discussed in the main text, a topological transition occurs at infinite temperature ($\beta = 0$) where the purity gap globally closes, highlighting the key role of the latter.}
    \label{fig:numericalResults1}
\end{figure}

We have used the above representation as input for the numerical evaluation of Eqs.~\eqref{eq:phasePathOrderedProduct} and~\eqref{eq:pathOrderedProduct} [which provide an exact reformulation of Eq.~\eqref{eq:EGP} for the EGP]. In Fig.~\ref{fig:numericalResults1}, we plot the difference $| \varphi_\text{E} - \varphi_\text{Z} |$ between the EGP and the Zak phase for fixed system parameters $t_1 - t_2, \Delta \neq 0$, over a wide range of $\beta$ including temperatures much larger than the Hamiltonian gap (of order $1$ for the chosen parameters). For all but the largest values of $T$, the data confirms the scaling $\sim 1/N^2$ predicted by perturbation theory [see Eq.~\eqref{eq:genericScalingCorrections}].

Next, we consider the EGP difference $\Delta \varphi_\text{E}$ accumulated over a closed path in parameter space $(t_1 - t_2, \Delta)$ encircling the origin (gapless point of $G_k = \beta H_k$). As argued in previous sections, we expect $\tfrac{1}{2\pi} \Delta \varphi_\text{E}$ to be an integer equal to the topologically quantized change (Chern number) $\tfrac{1}{2\pi} \Delta \varphi_\text{Z}$ of the Zak phase of the lowest band of $G_k = \beta H_k$ over the same parameter cycle. This equality must hold regardless of the system size $N$ and temperature $T$. The inset of Fig.~\ref{fig:numericalResults1} confirms this behavior: as the temperature is increased away from zero ($\beta$ decreased away from $\infty$), the difference $\tfrac{1}{2\pi} \Delta \varphi_\text{E}$ remains quantized for all system sizes $N$ accessible numerically. Moreover, its value indeed coincides with the value $\tfrac{1}{2\pi} \Delta \varphi_\text{Z} = 1$ corresponding to the quantized charge $\Delta P = 1$ that would be pumped through the system at $T = 0$. The topological quantization of $\tfrac{1}{2\pi} \Delta \varphi_\text{E}$ requires the spectral gap of $G_k = \beta H_k$ (the purity gap) to be finite all along the chosen cycle in parameter space (as required for ``purity adiabaticity''; see Sec.~\ref{subsec:purityAdiabaticity}). More importantly, the nontrivial value of $\tfrac{1}{2\pi} \Delta \varphi_\text{E}$ crucially depends on the existence of purity-gap-closing points encircled by the parameter cycle. In the thermal setting of interest here, the spectral gap of $G_k = \beta H_k$ can close either (i) via the closure of the energy gap $\Delta \epsilon$ of the underlying Hamiltonian $H_k$, or (ii) at infinite temperature where $\beta \to 0$. This leads to two possibilities for topological phase transitions. In the inset of Fig.~\ref{fig:numericalResults1}, the value of $\tfrac{1}{2\pi} \Delta \varphi_\text{E}$ computed for a range of negative to positive temperatures~\footnote{We consider negative temperatures for the sake of this argument.} illustrates these two possibilities: (i) when $| \beta | \neq 0$, the value of $\tfrac{1}{2\pi} \Delta \varphi_\text{E}$ is nontrivial because the parameter cycle that we consider encircles the purity-gap-closing point at the origin in parameter space $(t_1 - t_2, \Delta)$. It would be zero otherwise. (ii) When going from positive to negative temperatures, a topological transition occurs where $\tfrac{1}{2\pi} \Delta \varphi_\text{E}$ changes sign. Intuitively, the reason for this ``jump'' is that at positive/negative temperatures, the lower/upper band is predominantly occupied. At $\beta = 0$, occupation inversion occurs, and the sign of $\varphi_\text{E}$ changes. We note that negative temperatures strictly speaking do not correspond to an equilibrium situation anymore, although the generator of dynamics still is a Hamiltonian operator alone.

We now turn to a second illustrative example provided by the nonequilibrium analog of the Rice-Mele model introduced in Ref.~\cite{Linzner2016}. In that case, the relevant dynamics is governed by a gapped Liouvillian (Markovian quantum master equation) of the generic form
\begin{equation}
    \partial_t \rho = \sum_{r, s} \left( 2 L_{r, s} \rho L_{r, s}^\dagger - \{ L_{r, s}^\dagger L_{r, s}, \rho \} \right) \equiv \mathcal{L}(\rho),
    \label{eq:RMMasterEquation}
\end{equation}
where $\rho$ is the density matrix of the system and $L_{r, s}$ are so-called Lindblad operators. This type of time evolution can  be realized, e.g., in quantum systems with an engineered system-bath coupling. Provided that the bath energies are separated from those of the system by a large energy scale, Born-Markov and rotating-wave approximations become applicable, and lead to Lindblad master equations as above (see, e.g., Ref.~\cite{Bardyn2013} for details). Here we assume, for simplicity, that the right-hand side of Eq.~\eqref{eq:RMMasterEquation} does not contain a coherent contribution $\sim -i [\hat{H}, \rho]$.

The nonequilibrium Rice-Model model analog of interest is defined by the set of Lindblad operators
\begin{align}
    L_{r, 0} = \sqrt{1 + \epsilon} \Big[ &(1 - \lambda) \left( \hat{a}_{r, 0}^\dagger + \hat{a}_{r, 1} \right) \nonumber \\
    + &(1 + \lambda) \left( \hat{a}_{r, 0} - \hat{a}_{r, 1}^\dagger \right) \Big], \label{eq:Lindblad_A} \\
    L_{r, 1} = \sqrt{1 - \epsilon} \Big[ &(1 - \lambda) \left( \hat{a}_{r + 1, 0}^\dagger + \hat{a}_{r, 1} \right) \nonumber \\
    + &(1 + \lambda) \left( \hat{a}_{r + 1, 0} - \hat{a}_{r, 1}^\dagger \right) \Big], \label{eq:Lindblad_B}
\end{align}
where $L_{r, 0}$ and $L_{r, 1}$ act inside and between unit cells, respectively, and in this regard are analogous to the two hopping terms in Eq.~\eqref{eq:RiceMeleModel}. Referring to Ref.~\cite{Linzner2016} for details, we note that these operators are defined such that, on a timescale $\sim 1/\Delta_\text{d}$ set by the so-called ``damping gap'' $\Delta_\text{d}$ of the Liouvillian $\mathcal{L}$~\cite{Bardyn2013}, the dynamics drives an arbitrary initial Gaussian state to a specific stationary Gaussian state $\rho$ satisfying $\mathcal{L}(\rho) = 0$. The latter does not obey strict particle number conservation. Its mean particle number $\langle \hat{n} \rangle$, however, is stationary, and fluctuations $\langle \delta \hat{n} \rangle$ around it are intensive, i.e., $\langle \delta \hat{n} \rangle / \langle \hat{n} \rangle \to 0$ in the thermodynamic limit. For all practical purposes, the Gaussian state $\rho \sim e^{-G_k}$ can then be described by a number-conserving fictitious Hamiltonian $G_k$. Due to the specific form of the Lindblad operators, the matrix $G_k$ exhibits the same structure as the Hamiltonian matrix $H_k$ of the Rice-Mele model. Specifically, $G_k$ is given by Eq.~\eqref{Eq:RiceMele1stQuantized} with $\beta = 1$ (no notion of temperature here), $t_1 = \tfrac{1}{4}(1 + \epsilon)(\lambda^2 - 1)/(\lambda^2 + 1)$, $t_2 = \tfrac{1}{4}(1 - \epsilon)(\lambda^2 - 1)/(\lambda^2 + 1)$, and $\Delta = \lambda/(\lambda^2 + 1)$. Since $t_1 - t_2 \propto \epsilon$ and $\Delta \propto \lambda$, the real parameters $(\epsilon, \lambda)$ play a similar role as the parameters $(t_1 - t_2, \Delta)$ in the Rice-Mele model. In particular, the origin $\epsilon = \lambda = 0$ corresponds to a purity-gap-closing point. As in the thermal case, the EGP of the stationary state can be computed using Eqs.~\eqref{eq:phasePathOrderedProduct} and~\eqref{eq:pathOrderedProduct}.

\begin{figure}
    \includegraphics[width=\columnwidth]{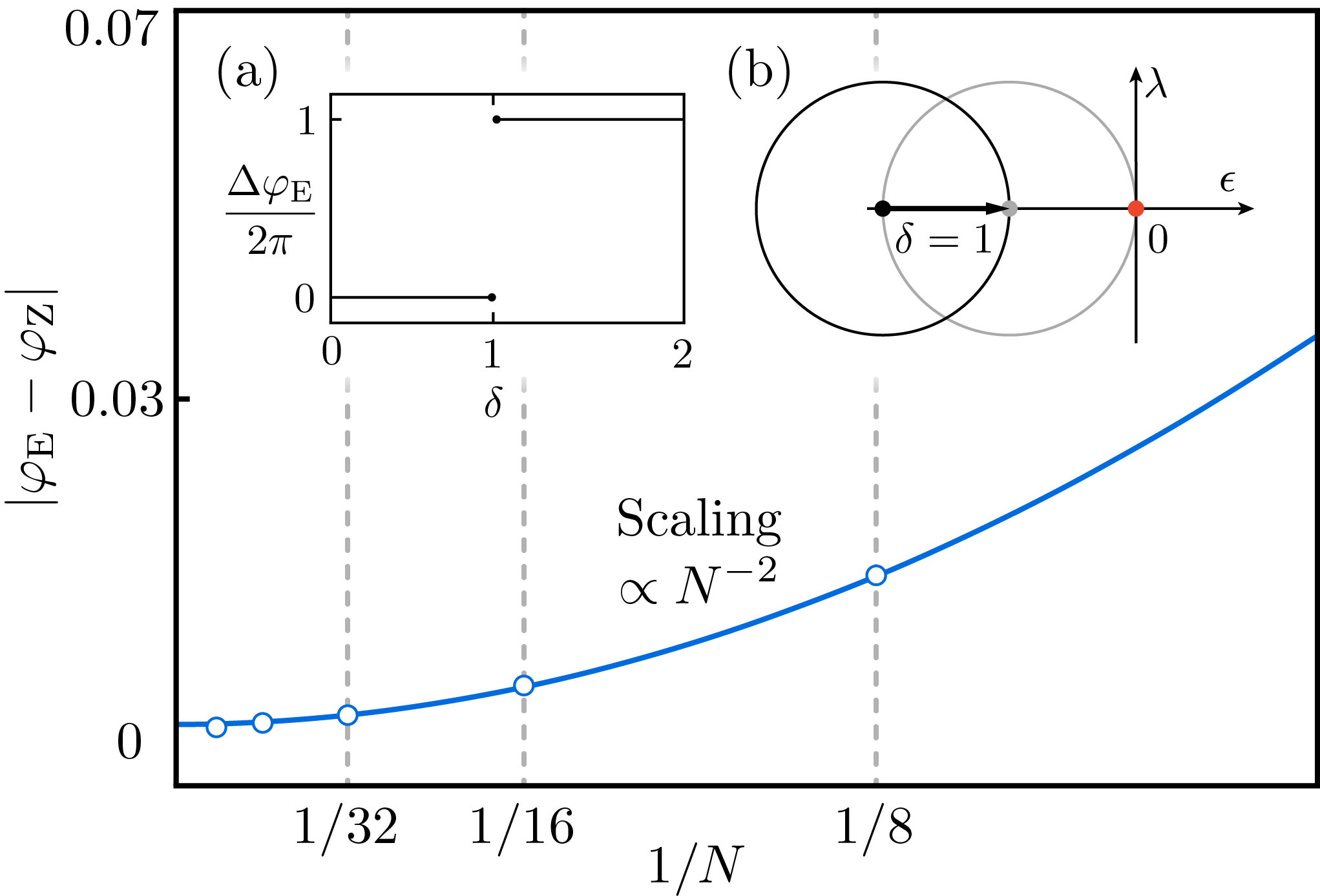} 
    \caption{Scaling of finite-size corrections $| \varphi_\text{E} - \varphi_\text{Z} |$ in the nonequilibrium analog of the Rice-Mele model introduced in Ref.~\cite{Linzner2016}, with parameters $\epsilon = -\sqrt{3}/4$ and $\lambda = -1/4$ in Eqs.~\eqref{eq:Lindblad_A} and~\eqref{eq:Lindblad_B}. As in the thermal case [Fig.~\ref{fig:numericalResults1}], the expected scaling behavior $\sim N^{-2}$ is verified. \textbf{Inset (a)}: Difference $\tfrac{1}{2\pi} \Delta \varphi_\text{E}$ accumulated per full cycle in $(\epsilon, \lambda)$ parameter space, as a function of a parameter $\delta$ controlling whether or not the purity-gap-closing point at $\epsilon = \lambda = 0$ is encircled [as illustrated in \textbf{inset (b)}, with gapless point shown in red]. As in the thermal case, quantization is observed irrespective of system size $N$, and a topological phase transition occurs at $\delta = 1$ where the purity-gap-closing point becomes encircled.}
    \label{fig:numericalResults2}
\end{figure}

In Fig.~\ref{fig:numericalResults2}, we show the computed difference $| \varphi_\text{E} - \varphi_\text{Z} |$ between the EGP and the Zak phase of the lowest purity band (lowest band of $G_k$). As in the thermal case, we verify that $\varphi_\text{E} \to \varphi_\text{Z}$ in the thermodynamic limit, with power-law scaling $\sim N^{-2}$. Purity-gap-closing points play the same key role here: in particular, we observe a topological transition where the EGP difference $\frac{1}{2\pi} \Delta \varphi_\text{E}$ accumulated over a complete cycle in $(\epsilon, \lambda)$ parameter space ``jumps'' from trivial ($0$) to nontrivial ($1$) when varying a parameter $\delta \equiv \delta(\epsilon, \lambda)$ controlling whether the purity-gap-closing point at $\epsilon = \lambda = 0$ is encircled [see insets of Fig.~\ref{fig:numericalResults2}]. In equilibrium thermal systems, purity-gap-closing \emph{points} necessarily coincide with points at which the gap of the system Hamiltonian closes, due to the tight correspondence $G_k = \beta H_k$. In nonequilibrium systems, in contrast, purity-gap-closing points need not coincide with points where the gap of the system Liouvillian (nonequilibrium analog of a Hamiltonian) closes. In fact, here, the gap of the Liouvillian (the damping gap) is given by $\Delta_\text{d} = 4 [1 + \lambda^2 (2 \epsilon^2 + \lambda^2 + 2 (\epsilon^2 - 1) \cos k)]^{1/2}$~\footnote{\unexpanded{We typically consider parameters with $|\lambda|, |\epsilon| < 1$, to avoid gap-closing points $|\lambda| = \pm 1$, $\epsilon = 0$.}}. Clearly, it does not close at the purity-gap-closing point $\epsilon = \lambda = 0$. The damping gap plays a similar role as a Hamiltonian gap, in the sense that it ensures the exponential decay of spatial correlations~\cite{Bardyn2013}. Therefore, here, the fact that it remains open at the point $\epsilon = \lambda = 0$ where the purity gap closes and a topological transition occurs exemplifies a remarkable possibility unique to nonequilibrium systems: the fact that topological transitions can occur without the appearance of divergent length or time scales (see Ref.~\cite{Bardyn2013} for a detailed discussion).

\section{Measurement of the EGP}
\label{sec:measurement}

As shown above, the EGP of a thermal or nonequilibrium state is not related to particle currents, which implies that it cannot be measured via particle transport. In the following, we propose an interferometric scheme to detect it, building on an idea by Sj\"oqvist \textit{et al.}~\cite{Sjoqvist2000}. To this end, we recall that $\varphi_\text{E} = \Im \ln \langle \hat{T} \rangle$ is nothing but the argument of the complex-valued observable $\langle \hat{T} \rangle$, i.e., $\varphi_\text{E} = \mathrm{arg} \langle \hat{T} \rangle$, where $\hat{T} = \exp (i \delta k \sum_i x_i \hat{a}_i^\dagger \hat{a}_i)$ [recall that $x_i$ denotes the position of fermions on site $i \equiv (r, s)$, with creation and annihilation operators $\hat{a}_i^\dagger, \hat{a}_i$].

\begin{figure}
    \includegraphics[width=\columnwidth]{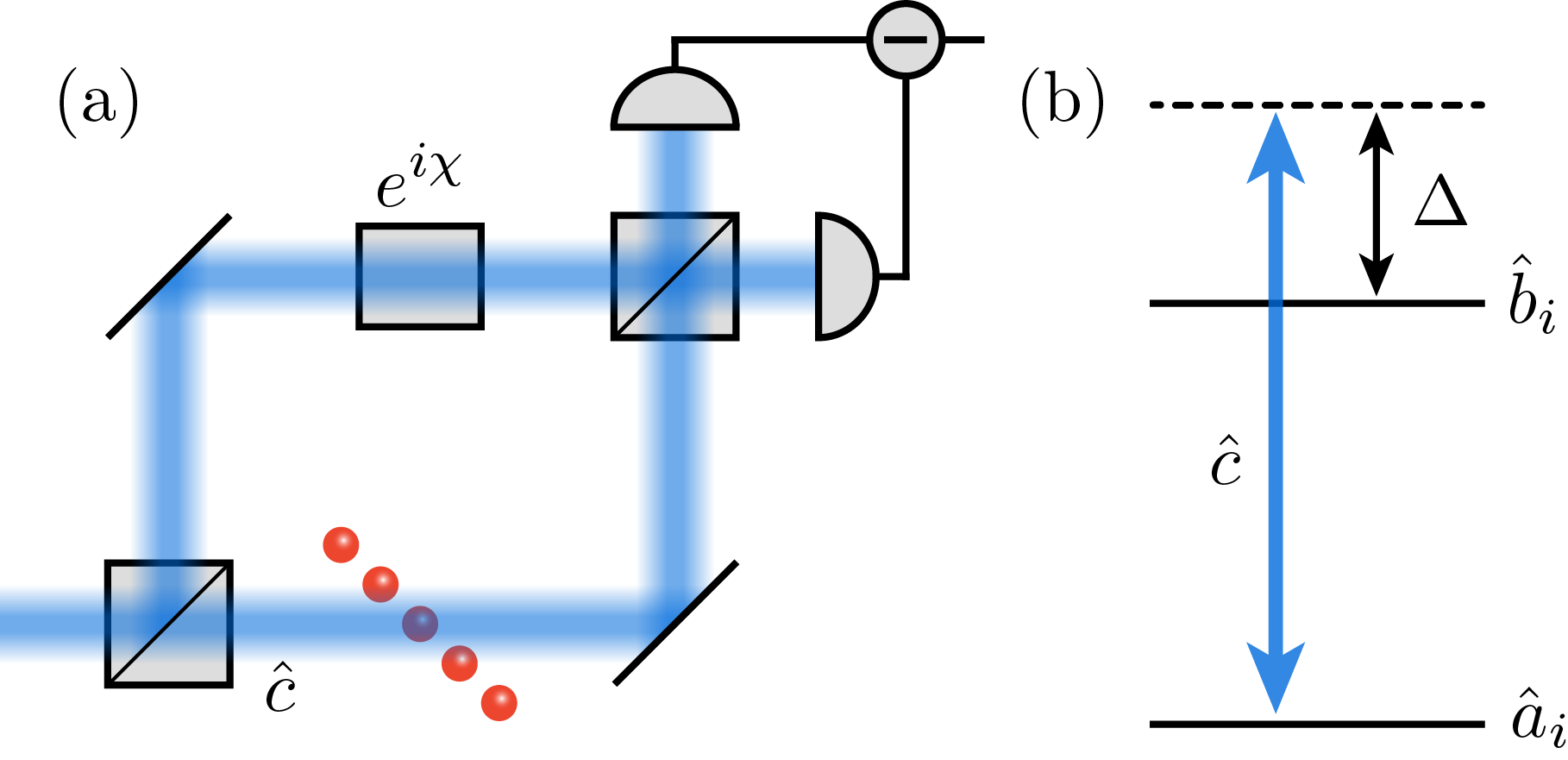}
    \caption{\textbf{(a)} Schematic setup for the interferometric measurement of the EGP. Photons are injected into one input and the intensity difference between the two outputs is detected. In the lower arm of the interferometer, each photon interacts with the 1D chain of fermions corresponding to the probed system. \textbf{(b)} Internal fermionic level scheme. On each site $i$, the photonic mode $\hat{c}$ couples ground and excited levels (with annihilation operators $\hat{a}_i$ and $\hat{b}_i$, respectively) in a far off-resonant way (large detuning $\Delta$).}
    \label{fig:interferometer}
\end{figure}

We consider a Mach-Zehnder interferometer whose lower arm contains the system to be probed [Fig.~\ref{fig:interferometer}(a)] --- fermions in a 1D lattice of length $L$ corresponding to a few tens of sites, as in typical setups with ultracold atoms. Photons moving along the two directions defined by the interferometer geometry can be represented by a two-state wavefunction with upper and lower components describing photons propagating in the ``vertical'' and ``horizontal'' directions, respectively. The action of mirrors and beam splitters is then described by unitary $2 \times 2$ matrices
\begin{equation}
    U_M = \left(\begin{array}{cc}
        0 & 1 \\
        1 & 0
    \end{array}\right), \quad
    U_B = \frac{1}{\sqrt{2}} \left(\begin{array}{cc}
        1 & i \\
        i & 1
    \end{array}\right),
\end{equation}
respectively. We assume that each lattice site $i$ consists of two internal fermionic levels (``ground'' and ``excited'') corresponding to annihilation operators $\hat{a}_i$ and $\hat{b}_i$, respectively. As illustrated in Fig.~\ref{fig:interferometer}b, fermions are coupled to a photonic mode with carrier frequency $\omega_0 = k_0 z$ (where $z$ is the position along the propagation path of interest), described by the annihilation operator $\hat{c}(z)$ satisfying the commutation relation $[\hat{c}(z), \hat{c}^\dagger(z^\prime)] = \delta(z - z^\prime)$. We assume that this mode couples the internal ground and excited states with coupling constant $g$ and a large detuning $\Delta$. When $\Delta$ is larger than all other relevant energy scales, internal excited states can be adiabatically eliminated, leading to the effective Hamiltonian
\begin{equation}
    H_\mathrm{eff} = \sum_j \frac{g^2}{\Delta} \vert f_j \vert^2\, \hat{a}_j^\dagger \hat{a}_j  \hat{c}^\dagger(z_j) \hat{c}(z_j),
    \label{eq:HScattering}
\end{equation}
where $f_j$ is the complex mode function of the photonic mode $\hat{c}(z)$ at site $j$. This Hamiltonian describes forward Brillouin scattering.

Next, we assume that the amplitude of the mode $f_j$ exhibits a spatial gradient along the axis of the probed system (realized, e.g., by a TEM$_{01}$ Gauss-Hermite mode), such that $\vert f_j \vert^2 \propto x_j$ (recall that $x_j$ is the position of the fermionic site $j$). Denoting $\vert f_j \vert^2 \equiv \eta \, x_j /L$, Eq.~\eqref{eq:HScattering} then describes the coupling between photons and the center-of-mass position operator $\hat{X} = \sum_j x_j \hat{a}_j^\dagger \hat{a}_j$ of the probed system, as desired. The propagation of the photonic mode along the lower arm of the interferometer is described by the equation
\begin{align}
    (\partial_z - i \omega_0) \hat{c}(z) & = -i \left[ \hat{c}(z), H_\mathrm{eff} \right] \\
    & = - i \sum_j \frac{g^2 \eta}{L \Delta} x_j \hat{a}_j^\dagger \hat{a}_j \delta(z - z_j) \hat{c}(z), \nonumber
\end{align}
(setting the speed of light $c = 1$), with solution
\begin{equation}
    \hat{c}(z) = e^{ik_0 z} \exp \left(-i \frac{g^2 \eta}{L \Delta} \hat{X} \right) \hat{c}(0),
\end{equation}
where $z$ lies beyond the region where photons interact with the probed system. By adjusting the detuning $\Delta$ such that $g^2 \eta/(L \Delta) = 2\pi/L \equiv \delta k$, photons in the lower arm of the interferometer pick up a phase proportional to the center of mass of fermions in the probed system --- described by the unitary transformation $\hat{T} = \exp(-i \delta k \hat{X})$, as desired.

We remark that photons additionally experience a momentum ``kick'' in the direction of the lattice of the probed system due to the spatial gradient in their mode function. The kick imparted to each photon, however, is less than their initial momentum $k_0$. This effect thus only leads to a small diffraction which we neglect here, for simplicity (though it should to be taken into account when designing an actual experiment). We note that the prefactor $g^2 \eta/(\Delta L)$ can be increased, at fixed detuning $\Delta$, by adding a build-up cavity to the lower arm of the interferometer to make photons bounce back and forth through the probed system before propagating further --- thereby enhancing the effective interaction between photons and fermions.

The total unitary matrix describing the propagation of a photon through the interferometer reads
\begin{equation}
    \mathbf{U} = \mathbf{U}_B \mathbf{U}_M \mathbf{U}_\text{int} \mathbf{U}_M\mathbf{U}_B,
\end{equation}
where $\mathbf{U}_B = U_B \otimes \mathbb{1}$ and $\mathbf{U}_M = U_M \otimes \mathbb{1}$ ($\mathbb{1}$ being the unit operator in the Hilbert space of the fermions), and
\begin{equation}
    \mathbf{U}_\text{int} = \left(\begin{array}{cc}
        0 & 0 \\
        0 & 1
    \end{array}\right) \otimes \, \hat{T} + \left(\begin{array}{cc}
        e^{i\chi} & 0 \\
        0 & 0
    \end{array}\right) \otimes \, \mathbb{1},
\end{equation}
where $\chi$ is a tunable phase in the upper arm of the interferometer [Fig.~\ref{fig:interferometer}(a)]. Overall, the interferometer transforms the input state $\varrho_\text{in} = \rho_\text{ph} \otimes \rho$, where $\rho_\text{ph}$ is the input state of photons and $\rho$ is the mixed state of the probed system, into $\mathbf{U} \varrho_\text{in} \mathbf{U}^\dagger$. The resulting intensity at the two output ports (``$+$'' and ``$-$'') is then given by
\begin{equation}
    \hat{n}_\mathrm{out}^{\pm} = \frac{1}{2} \left( 1 \pm \vert \langle \hat{T} \rangle \vert \cos \bigl[ \chi - \mathrm{arg} \langle \hat{T} \rangle \bigr] \right) \hat{n}_\mathrm{in},
\end{equation}
where $\hat{n}$ denotes photon number operators and the expectation value $\langle \hat{T} \rangle$ is only over the fermionic state of the probed system. Therefore, in the above setup, photons injected into the lower arm of the interferometer pick up a phase which crucially corresponds to the EGP $\varphi_\text{E} = \mathrm{arg} \langle \hat{T} \rangle$. Most importantly, this phase can be measured by monitoring directly the intensity difference $\Delta \hat{n} \equiv \hat{n}_\text{out}^+ - \hat{n}_\text{out}^- = \vert \langle \hat{T} \rangle \vert \cos \bigl [\chi - \mathrm{arg} \langle \hat{T} \rangle \bigr] \hat{n}_\mathrm{in}$ between outputs (balanced detection) as a function of the reference phase $\chi$ set in the upper arm. To extract the accumulated phase difference $\Delta \varphi_\text{E} = \oint d\phi \, \partial_\phi \varphi_\text{E}$, one can then (i) repeat the experiment for multiple parameter values $\phi$ along a cyclic path, and (ii) follow the procedure discussed in Sec.~\ref{subsec:purityAdiabaticity} to extract the exact integer value of $\Delta \varphi_\text{E}$ in a way that is gauge-invariant and, most importantly, robust against (small) measurement errors on $\varphi_\text{E}$.

The visibility of the EGP signal resulting from the above balanced detection scheme is unity. Since $| \langle \hat{T} \rangle |$ is typically small, however, the characteristic number of output photons per input photon is small and one may have to accumulate photons for a longer time to reach the desired sensitivity in the measurement of $\varphi_\text{E}$. The minimum detectable phase is set by shot noise $\varphi_\text{E} \bigr \vert_\mathrm{min} = \Delta \varphi_\mathrm{shot} \sim 1/\sqrt{P_\mathrm{out} \tau}$, where $P_\mathrm{out}$ is the maximum output flux of photons per unit time and $\tau$ is the overall measurement time.

In the illustrative two-band model with purity spectrum $\pm \beta_k$ examined above, one finds $| \langle \hat{T} \rangle | = \exp[- \tfrac{N}{2\pi} \int dk \ln(1 + e^{-\beta_k}) + \mathcal{O}(1/N^2)] \approx \exp(- \tfrac{N}{2\pi} e^{-\Delta \beta / 2})$ (with an additional factor $\sin\theta$), where $\Delta \beta = 2 \min_k \beta_k$ is the purity gap~\footnote{\unexpanded{Note that the real factor $\ln \det [1 - f(G)]$ neglected in Eq.~\eqref{eq:phasePathOrderedProduct} must be taken into account in $| \langle \hat{T} \rangle |$.}}. Note that also $\vert\langle T\rangle\vert\approx 1$ can be achieved e.g., in a mesoscopic thermal system of $N \approx 50$ sites with a purity gap $\Delta \beta = \Delta \epsilon / T \approx 5$ (where $\Delta \epsilon$ is the energy gap of the underlying Hamiltonian).

We finally comment on the effects of boundary conditions: although we have always considered a 1D lattice system of fermions with \emph{periodic} boundary conditions, for pedagogical purposes, we emphasize that the EGP is a bulk quantity (reducing to the polarization, in the zero-temperature limit) which is, hence, essentially unaffected by boundary conditions. In particular, \emph{open} boundary conditions can also be used, as implicitly assumed in the above measurement setup. In that case, correlations $\langle \hat{a}^\dagger_i \hat{a}_j \rangle = [f(G)]_{ij}$ between sites $i$ and $j$ located at opposite ends of the system --- corresponding to the corner elements of the matrix $f(G)$ --- are essentially removed~\footnote{Correlations correspond to the state defined in Eq.~\eqref{eq:GaussianDensityMatrix}, which can be seen as the thermal density matrix of a fictitious Hamiltonian $G$ (with temperature $\beta = 1$). Accordingly, correlations must decay exponentially with system size over a length scale proportional to the inverse spectral gap of $G$ (the purity gap)~\cite{Hastings2004}. As in the rest of this work, we assume that the purity gap is finite.}, and the matrix $f(G)(T - 1)$ appearing in Eq.~\eqref{eq:EGPFormal} for the EGP (namely, $\varphi_\text{E} = \Im \ln \det [\mathbb{1} + f(G) (T - \mathbb{1})]$) is similarly quasi-diagonal in position space, with \emph{no corner elements}. Since $\det [\mathbb{1} + f(G) (T - \mathbb{1})] = \exp \Tr \ln [\mathbb{1} + f(G) (T - \mathbb{1})]$, elements of $f(G) (T - \mathbb{1})$ contribute to $\varphi_\text{E}$ to order $m$ (in a series expansion of the logarithm) whenever they contribute to diagonal elements in $[f(G) (T - \mathbb{1})]^m$. Here, quasi-diagonal elements in $f(G)(T - 1)$ contribute at order $\sim 1$, while corner elements contribute at order $\sim N$ (where $N$ is the total number of sites in the system). As a result, corner elements determining boundary conditions lead to corrections to the EGP that are \emph{exponentially} small with system size, as we have verified numerically. In particular, the plots presented in Sec.~\ref{sec:examples} are visually unchanged when considering open instead of periodic boundary conditions.

\section{Conclusions and outlook}
\label{sec:conclusions}

We have shown that density matrices describing mixed fermionic Gaussian states in one dimension encode topological information in a way that enables a direct interpretation in terms of a physical observable. The connection to observables is provided by the ensemble geometric phase (EGP), which is defined for arbitrary density matrices --- equilibrium and non-equilibrium states alike. The EGP is constructed from the expectation value of a many-particle operator: the operator of translations in momentum space by the smallest possible step $\delta k = 2\pi/L$, which crucially equips us with a geometric notion of parallel transport for state vectors in that space. The mechanism underlying the robustness of the EGP as a geometric phase for mixed states is based on the statistical selection of the most strongly occupied Bloch band, i.e., typically the lowest one. This selection is a many-body effect related to the presence of $N$ modes in each Bloch band, leading to an effective physical purification of the selected Bloch band in the thermodynamic limit $N \to \infty$.

Although the price to pay to see fingerprints of topology in mixed states is the many-body character of the EGP, we have shown that the latter can be detected in interferometric measurements, e.g., in current setups based on cold atomic gases. These results have two important physical implications: conceptually, they demonstrate that topological phase transitions persist to finite temperatures, or, more generally, in mixed quantum states. More precisely, the degeneracy in the spectrum of the density matrix, measured by the purity-gap closing, is associated with a singularity in the EGP, i.e., with a jump in the accumulated EGP differences upon enclosing such a point in parameter space. Practically, the EGP provides a viable \emph{in situ} alternative to detecting topological order in fermionic systems of ultracold atoms in low-temperature states and at finite density. So far, the Zak phase has been determined in the single-particle limit only, where particle statistics is irrelevant, by propagating a test particle through an otherwise empty band structure~\cite{Bloch2013} (see Ref.~\cite{Jotzu2014} for a related strategy in a two-dimensional system). Cooling fermions to extremely low temperatures to access topological ground-state properties remains an outstanding challenge at finite fermion density. Due to its robustness towards finite temperatures, however, the EGP examined here could be used as a direct detection tool in such systems.

Our construction based on a unitary translation operator may lend itself to generalizations to interacting systems~\cite{Shapourian2017, Shiozaki2017} in mixed states. Here we have focused on noninteracting, translation-invariant systems, and the crystal momentum may have seemed to play a very fundamental role. We emphasize, however, that the truly relevant object is the many-body translation operator $\hat{T} = e^{i \delta k \hat{X}}$ used to define the EGP. This operator describes a shift $\delta k = 2\pi / L$ of the \emph{physical} momentum of all particles, irrespective of interactions or disorder. It acts as a canonical transformation $\hat{p}_j \to \hat{T}^\dagger \hat{p}_j \hat{T} = \hat{p}_j + \delta k$ (where $\hat{p}_j$ is the momentum operator of particle $j$), which can conveniently be seen as the insertion of a magnetic flux $\Phi = 2\pi$ through the periodic system~\cite{ThoulessNiu}. Therefore, in the general case where the crystal momentum $k$ is not a good quantum number, $\Phi$ simply replaces $k$ as the relevant parameter for state vectors (as in many-body generalizations of the Zak phase and of the Chern number~\cite{ThoulessNiu}). In that case the operator $\hat{T}$ defines, for vectors $\ket{\psi_\Phi}$, a similar notion of parallel transport in $\Phi$ space as it does in $k$ space for vectors $\ket{\psi_k}$ in noninteracting systems with translation invariance. As a result, we expect the gauge-reduction mechanism identified in this work to hold in interacting or disordered systems, as long as the density matrix $\rho$ of the system (or, more precisely, the corresponding ``fictitious Hamiltonian'' $-\ln \rho$) has a gapped ground state. In particular, we expect the EGP to remain nontrivial when weak interactions and/or disorder preserving the purity gap are added to the examples examined in Sec.~\ref{sec:examples}.

Beyond interacting and disordered systems, several directions will be exciting to explore: First, the gauge-reduction mechanism identified here seems very generic. In particular, we anticipate that one would obtain other well-defined geometric phases for mixed states by replacing the operator $\hat{T}$ of translations in momentum space by other unitary operators, generating translations in a different space (at the expense, however, of possibly losing the direct connection to physical observables which is a key feature of the EGP). Second, and more broadly, an intriguing direction for future research will be to examine whether other types of gauge structures can be extracted from many-body correlators~\cite{Shapourian2017, Shiozaki2017}, to construct topological classifications of mixed states. Extensions to systems with more than one spatial dimension will also be interesting to explore. Third, it will be exciting to ask whether the gauge-reduction mechanism identified in fermionic systems here can also play a role in bosonic ones. While bosons at equilibrium and low temperature tend to condense with a strong occupation of low-momentum modes, experiments based, e.g., on ultracold atoms in modulated potentials~\cite{Aidelsburger2015} could lead to topological signatures in the EGP. Exciton polaritons are also potential candidates, as they routinely produce driven open quantum states with occupation properties reminiscent of fermionic systems~\cite{Baboux2017}.

Finally, our work may pave the way towards other probes of mixed-state topology. One promising candidate are Loschmidt amplitudes --- another many-body observable related to the expectation value of a unitary matrix, namely, the time-evolution operator of Hamiltonian quantum dynamics. Cases which could perhaps be related to our mechanism were recently pointed out in Refs.~\cite{Dutta2017, Budich2017}. Fingerprints of bulk topological properties at the edges of insulators and superconductors have been shown to persist at finite temperatures~\cite{Quelle2016, Kempkes2016}, and establishing a connection to the results presented here provides another challenge for future research.

\section*{Acknowledgements}

We would like to thank D. Linzner for invaluable input and support, and J. C. Budich and M. Heyl for useful discussions. C.-E.~B. gratefully acknowledges support from the DQMP at the University of Geneva and from the Swiss National Science Foundation under Division II. S.~D. acknowledges support by the German Research Foundation (DFG) through the Institutional Strategy of the University of Cologne within the German Excellence Initiative (ZUK 81), as well as support by the European Research Council via ERC Grant Agreement n. 647434 (DOQS). L. W. and M. F. acknowledge support by the German Research Foundation (DFG) through the SFB TRR 185, and A.~A. and S.~D. (DFG) through the CRC 183 (project B02). C.-E.~B., M.~F. and S.~D. would like to extend specials thanks to the KITP at UCSB for hospitality. This research was supported in part by the National Science Foundation under Grant No. NSF PHY-1125915.


\appendix
\section*{Appendices} 

\section{EGP vs. adiabatic pump currents}
\label{sec:EGPvsAdiabaticPumpCurrents}

In this appendix, we discuss the connection between the EGP and the current flow induced in an adiabatic pump protocol. The main conclusion will be that while $\tfrac{1}{2\pi} \Delta \varphi_\text{E}$ is quantized and physically observable, it does not correspond, for mixed states, to a quantized charge transfer. This is surprising inasmuch as the EGP does coincide with the adiabatic current in the zero-temperature (ground-state) limit. In the following, we start by reviewing the connection between Resta's formula for the ground-state polarization of periodic quantum systems and the current flow.

\subsection{Resta polarization and current flow}
\label{subsec:}

Resta formula for the ground-state polarization $P$ of a periodic quantum system [Eq.~\eqref{eq:RestaPolarization}] is constructed in such a way that differential changes $\partial_t P \equiv \dot{P}$ induced, e.g., by an external parameter are equal to the physical current $I$ (as they should). Here, we review how the connection between Eq.~\eqref{eq:RestaPolarization} and the current explicitly arises, in line with Resta's arguments in Ref.~\cite{Resta1998}.

The starting point of the construction is the observation that $\hat{T} | \psi_0 \rangle = e^{i \delta k \hat{X}} | \psi_0 \rangle$ corresponds, to first order in $\delta k = 2\pi/L$, to the ground state of the momentum-shifted Hamiltonian $\hat{H}(q = -\delta k) \equiv \hat{T} \hat{H} \hat{T}^\dagger \simeq \hat{H} + i \delta k [\hat{X}, \hat{H}] = \hat{H} - \delta k \hat{I}$, where $\hat{I} \equiv \partial_q \hat{H}(q)$ is the usual current operator. The action of $\hat{T}$ can be understood as a momentum shift of all particle momenta $\hat{p}_j \to \hat{T} \hat{p}_j \hat{T}^\dagger = \hat{p}_j - \delta k$. Equivalently, one can see $\hat{T} \hat{H} \hat{T}^\dagger$ as the Hamiltonian describing the periodic 1D system of interest after the adiabatic insertion of a single quantum of magnetic flux through the latter. To first order in $\delta k$, we obtain
\begin{equation}
    \hat{T} | \psi_0 \rangle = e^{i \varphi_\text{Z}} \left( | \psi_0 \rangle - \delta k \sum_{n \neq 0} | \psi_n \rangle \frac{\langle \psi_n | \hat{I} | \psi_0 \rangle}{E_0 - E_n} \right),
    \label{eq:GroundStateFlux}
\end{equation}
where $\varphi_\text{Z}$ is the Zak phase accumulated during the insertion of the flux quantum. In accordance with Resta's formula [Eq.~\eqref{eq:RestaPolarization}], the geometric phase $\varphi_\text{Z}$ determines the instantaneous value of the polarization, namely, $P = \tfrac{1}{2\pi} \Im \ln \langle \psi_0 | \hat{T} | \psi_0 \rangle = \tfrac{1}{2\pi} \Im \ln \langle \psi_0 | e^{i \varphi_\text{Z}} | \psi_0 \rangle = \varphi_\text{Z}/(2\pi)$. The Zak phase, however, crucially does not enter the expression of the current flow, which involves excitations out of the ground state. Indeed, starting from Resta's formula and using Eq.~\eqref{eq:GroundStateFlux}, the derivative $\dot{P}$ takes the form
\begin{align}
    \dot{P} & = \frac{1}{2\pi} \Im \frac{\langle \dot{\psi}_0 | \hat{T} | \psi_0 \rangle + \langle \psi_0 | \hat{T} | \dot{\psi}_0 \rangle}{\langle \psi_0 | \hat{T} | \psi_0 \rangle} \nonumber \\
    & = -\frac{\delta k}{\pi} \Im \sum_{n \neq 0} \frac{\langle \dot{\psi}_0 | \psi_n \rangle \langle \psi_n | \hat{I} | \psi_0 \rangle}{E_0 - E_n}.
    \label{eq:GroundStatePPumpCurrent}
\end{align}
This expression was identified in Ref.~\cite{ThoulessNiu} as the pump current flowing in response to the adiabatic variation of external parameters. For the sake of completeness, we review this connection in the next subsection in a manner somewhat different from the original derivation, tailored to the present discussion. Readers wishing to proceed directly to the generalization to mixed states may skip this discussion.

\subsection{For ground states, the time derivative of the EGP [Eq.~\eqref{eq:GroundStatePPumpCurrent}] coincides with the pump current}
\label{sec:ZakSpectralCurrent}

Let $\ket{\psi_0(t)} \equiv \ket{\psi_0(\phi)}$ be the instantaneous ground state at a value $\phi \equiv \phi(t)$ of a varied external parameter (such that $\hat{H}(\phi) \ket{\psi_0(\phi)} = E_0(\phi) \ket{\psi_0(\phi)}$). We assume that $\ket{\psi_0(\phi)}$ is current-free, i.e., $\bra{\psi_0(t)} \hat{I} \ket{\psi_0(t)} = 0$. When the parameter $\phi$ is adiabatically varied, the time evolution of the system initially prepared in its ground state is described by the time-dependent Schr\"odinger equation $i \partial_t \ket{\psi} = \hat{H} \ket{\psi}$ (omitting explicit time dependences, for simplicity). A convenient ansatz consists in writing $\ket{\psi} \equiv e^{-i (E_0 t - \varphi)} \ket{\psi_0} + \ket{\delta \psi} \equiv \ket{\psi'_0} + \ket{\delta \psi}$, with dynamical phase $E_0 t$, geometric phase $\varphi \equiv \varphi(t)$, and out-of-ground-state excitations described by $\ket{\delta\psi}$. Since the parameter $\phi$ is varied with frequency $\omega$ smaller than the excitation gap $\Delta \equiv \min_n(E_n - E_0)$ (adiabacity condition), $\ket{\delta \psi}$ will be small, but nevertheless important, as it is the wavefunction component responsible for the current flow, $\langle \hat{I} \rangle(t) \simeq \bra{\psi'_0(t)} \hat{I} \ket{\delta \psi(t)} + (\text{c.c.})$. Substitution of the ansatz into the Schr\"odinger equation yields
\begin{equation}
    e^{-i (E_0 t - \varphi)} (\dot{\varphi} \ket{\psi_0} + i | \dot{\psi_0} \rangle) + (i \partial_t - H) \ket{\delta \psi} = 0.
    \label{eq:SchroedingerEqPump}
\end{equation}
To make progress, we expand $\ket{\delta \psi} \equiv \sum_{n \neq 0} c_n e^{-i E_n t} \ket{\psi_n}$ in terms of the instantaneous excited eigenstates of the Hamiltonian, where $c_n$ are the time-dependent coefficients to be solved for, with initial condition $c_n(0) = 0$. In view of the expected smallness $c_n\sim \omega$, the temporal variation $| \dot{\psi}_n \rangle \sim \omega$ can be neglected as higher order, such that $(i \partial_t - H) \ket{\delta \psi} \simeq \sum_n i\dot{c}_n e^{-i E_n t} \ket{\psi_n}$. Substitution of this expression into Eq.~\eqref{eq:SchroedingerEqPump} followed by a projection onto $\ket{\psi_0}$ then identifies $\dot{\varphi} = i \langle \psi_0 | \dot{\psi}_0 \rangle$ as the Berry connection of the ground state (as expected). Projection onto excited states $\ket{\psi_n}$, on the other hand, leads to the equation $\dot{c}_n = e^{i (E_n - E_0) t} \langle \psi_n | \dot{\psi}_0 \rangle$, where we have neglected a Berry-phase factor as higher order. Likewise neglecting the slow variation of the matrix elements $\langle \psi_n | \psi_0 \rangle$ in comparison to the dynamical factor $e^{i (E_n - E_0) t}$, we obtain $c_n \simeq i (E_n - E_0)^{-1} \langle \psi_n | \dot{\psi}_0 \rangle [1 - e^{i (E_n - E_0) t}]$. Inserting this solution (neglecting rapidly oscillatory factors) into the spectral decomposition of the current expectation value $\langle \hat{I} \rangle(t) \simeq \bra{\psi'_0(t)} \hat{I} \ket{\delta \psi(t)} + (\text{c.c.})$ finally leads to Eq.~\eqref{eq:GroundStatePPumpCurrent}.

For completeness, we note that the current accumulated during a periodic pump cycle, $\Delta P = \oint dt \, \partial_t P = \oint d\phi \partial_\phi P$, is integer quantized. In Ref.~\cite{ThoulessNiu}, this quantization was established in a three-step argument: first, the above derivation was generalized to include the presence of a general magnetic flux $\Phi \in [0, 2\pi/L] = [0, \delta k]$ threading the system. Second, it was shown that the current expectation value is the same at any value of the flux, i.e., $\langle \hat{I} \rangle(\Phi) \simeq \langle \hat{I} \rangle(0)$ (up to corrections that vanish in the thermodynamic limit), such that the relevant current can be expressed as the average $\langle \hat{I} \rangle \equiv \delta k^{-1} \oint d\Phi \, \langle \hat{I} \rangle(\Phi)$. Third, it was argued that the presence of the flux leads to a modification of the Hamiltonian $\hat{H} \to \hat{H} + \Phi \hat{I}$, such that perturbation theory similar to the one outlined above yields $(E_0 - E_n)^{-1} \langle \psi_0 | \hat{I} | \psi_n \rangle = \langle \psi_0 | \partial_\Phi \psi_n \rangle$ to first order in $\delta k$ (where wavefunctions now depend on $\phi$ and $\Phi$). Using the additional identity $| \dot{\psi}_0 \rangle = \dot{\phi} | \partial_\phi \psi \rangle$, the charge transported per adiabatic cycle becomes
\begin{equation}
    \Delta P = \frac{i}{2\pi} \iint d\Phi d\phi \left( \langle \partial_\phi \psi_0 |\partial_\Phi \psi_0 \rangle - \langle \partial_\Phi \psi_0 | \partial_\phi \psi_0 \rangle \right).
\end{equation}
We recognize here the standard expression for the first Chern number of the Berry connection $A_j = i \langle \psi_0 | \partial_j \psi_0\rangle$ (with $j = \phi, \Phi$), which shows that $\Delta P$ is indeed a topologically quantized integer.

\subsection{For generic states, the time derivative of the EGP differs from the pump current}
\label{subsec:EGPSpectralCurrent}

We have shown in the main text that the EGP is a physical observable and that its integrated change over a complete cycle in parameter space is integer quantized. We have also argued that the EGP reduces to the Zak phase ($2\pi$ times the polarization) in cases where the density matrix reduces to a ground-state projector. Nonetheless, its topological quantization cannot be interpreted in terms of current flow, as we demonstrate now.

For the sake of concreteness, consider a thermal density matrix with general spectral decomposition $\rho = \sum_n p_n | \psi_n \rangle \langle \psi_n |$, where $p_n = e^{-\beta E_n}/\mathcal{Z}$ and $\mathcal{Z} = \sum_n e^{-\beta E_n}$. When introducing a time-dependent parameter $\phi \equiv \phi(t)$, both $E_n(\phi)$ and the states $| \psi_n(\phi) \rangle$ become time dependent. The time derivative of the EGP defined by Eq.~\eqref{eq:EGP} then becomes
\begin{align*}
    & \partial_t \varphi_\text{E} = \frac{1}{2\pi} \Im \frac{1}{\sum_n p_n \langle \psi_n | \hat{T} | \psi_n \rangle} \times \cr
    & \quad \times \sum_n \left( \dot{p}_n \langle \psi_n | \hat{T} | \psi_n \rangle + p_n \langle \dot{\psi}_n | \hat{T} | \psi_n \rangle + p_n \langle \psi_n | \hat{T} | \dot{\psi}_n \rangle \right).
\end{align*}
This expression is nonlinear in the occupation numbers $p_n$ and, hence, does not lend itself to analytical simplifications. Assuming that parameter changes are slow enough for an adiabacity principle to hold for individual states with weight $p_n$, the matrix elements appearing in this expression will carry geometric phases [see Eq.~\eqref{eq:GroundStateFlux}] which generally do not cancel out [except in the specific case of ground states $\rho = \ket{\psi_0} \bra{\psi_0}$ where the above expression reduces to Eq.~\eqref{eq:GroundStatePPumpCurrent}]. Even for time-independent states $(\dot{p}_n = 0)$, the presence of a nontrivial sum of operator expectation values in the denominator makes the above expression formally different from linear-response expectation values describing current flows.

\section{Perturbative corrections to the EGP}
\label{sec:perturbativeCorrections}

\begin{widetext}
In this appendix, we provide additional details regarding the second-order perturbative expansion of the EGP [Eq.~\eqref{eq:EGPPerturbativeExpansion}] and the scaling of second-order corrections [Eq.~\eqref{eq:genericScalingCorrections}] in the same context as in Sec.~\ref{subsec:twoBandExample}, i.e., in an illustrative two-band model with purity spectrum $\pm \beta_k$. For convenience, we focus on the limit of large system sizes $N$ where sums over momenta can be approximated as continuous integrals. Second-order corrections then read
\begin{align}
    \Delta(N) & = \Im \Tr G V^{(2)} + \tfrac{1}{2} \Im \Tr (G V^{(1)})^2, \nonumber \\
    & V^{(1)} = \int_0^{2\pi} dk \left( W^{-}_{2\pi, k} V_k^{-+} W^{+}_{k, 0} + W^{+}_{2\pi, k} V_k^{+-} W^{-}_{k, 0} \right), \nonumber \\
    & V^{(2)} = \int_0^{2\pi} dk \int_0^k dk' \left(W^{-}_{2\pi, k} V_k^{-+} W^{+}_{k, k'} V^{+-}_{k'} W^{-}_{k', 0} + W^{+}_{2\pi, k} V_k^{+-} W^{-}_{k, k'} V^{-+}_{k'} W^{+}_{k', 0} \right), \nonumber \\
    & G = P^{+} - (W^{-}_{2\pi, 0})^{-1} P^{-}.
\end{align}
Here, upper and lower bands are identified by ``$+$'' and ``$-$'' indices, respectively, with Bloch states $| u_k^\pm \rangle$. The operators $W^\pm_{k_1, k_2} = \exp[ \int_{k_1}^{k_2} dk \, (i A_k^{\pm} \pm \delta k^{-1} \beta_k)] \, \mathbb{1}$ describe the ``evolution'' in individual bands under the influence of the Berry connection $A_k^\pm = \mathcal{A}_k^0 \pm \mathcal{A}_k^3 = i \langle u_k^\pm | \partial_k u_k^\pm \rangle$ and the weight factors $\pm \beta_k$. The operators $V_k^{-+} = (\mathcal{A}_k^1 - i \mathcal{A}_k^2) \sigma^+$ and $V_k^{+-} = (\mathcal{A}_k^1+ i \mathcal{A}_k^2) \sigma^-$, on the other hand, are ``jump operators'' causing transitions between bands [with $\sigma^\pm \equiv (\sigma_1 \pm i \sigma_2)/2$]. The operator $G$ is the ``Green's function'' defined in Eq.~\eqref{eq:EGPPerturbativeExpansion}, which acts trivially in the upper band, and via the inverse evolution operator in the lower band [where $P^{\pm} = (\sigma_0 \mp \sigma_3)/2$ projects onto individual bands]. Using the above expressions, the second-order corrections can be cast in the form
\begin{align}
    \Delta(N) = \Im \Tr \int_0^{2\pi} dk \int_0^k dk' \Big[ & - (W^{-}_{k, k'})^{-1} V^{-+}_k W^{+}_{k, k'} V^{+-}_{k'} + W^{+}_{2\pi, 0} (W^{+}_{k, k'})^{-1} V^{+-}_k W^{-}_{k, k'} V^{-+}_{k'} \nonumber \\
    & - (W^{-}_{2\pi, k'})^{-1} V^{-+}_k W^{+}_{2\pi, 0} W^{+}_{k, k'} V^{+-}_{k'} - W^{+}_{2\pi, 0} (W^{+}_{k, k'})^{-1} V^{+-}_k W^{-}_{k, k'} V^{-+}_{k'} \Big],
    \label{eq:secondOrderCorrections4Terms}
\end{align}
where the first and second lines are the contributions of $V^{(2)}$ and $(V^{(1)})^2$, respectively, and where we have used the properties $(W^\pm_{k_1, k_2})^{-1} = W^\pm_{k_2, k_1}$ and $W^\pm_{k_1, k_2} W^\pm_{k_2, k_3} = W^\pm_{k_1, k_3}$. In equation~\eqref{eq:secondOrderCorrections4Terms}, the second and fourth terms cancel out, and the third term is massively suppressed due to the presence of $W^{+}_{2\pi, 0} \sim \exp(- \bar{\beta} N)$ with $\bar{\beta} \equiv (2\pi)^{-1} \int_0^{2\pi} dk \, \beta_k$. We are thus left with the first term, and noting that $(W^{-}_{k, k'})^{-1} W^{+}_{k, k'} = \exp[ 2 \int_{k'}^k dq (i \mathcal{A}_q^3 - \delta k^{-1} \beta_k)]$, we obtain
\begin{equation}
    \Delta(N) = - \Im \int_0^{2\pi} dk \int_0^k dk' (\mathcal{A}_k^1 - i \mathcal{A}_k^2) (\mathcal{A}_{k'}^2 + i \mathcal{A}_{k'}^2) e^{2 \int_{k'}^k dq (i \mathcal{A}_q^3 - \delta k^{-1} \beta_q)}.
\end{equation}
This expression makes the essence of the EGP gauge-reduction mechanism manifest: due to the global presence of weight factors $\sim \exp(- \delta k^{-1} \beta_k) \sim \exp(- N \beta_k)$, excursions from the ground state to the higher band are exponentially costly and effectively confined to short intervals $\sim (N \Delta \beta)^{-1}$ in the Brillouin zone, where $\Delta \beta = 2 \min_k{\beta_k}$ is the purity gap. A quantitative estimate of the suppression factor may be obtained by noting that, for $k = k'$, the integrand is real. A straightforward Taylor expansion to first order in $k - k'$ then yields the leading-order contribution
\begin{align}
    \Delta(N) & \simeq \int_0^{2\pi} dk \int_0^k dk' (k - k') e^{-2 \delta k^{-1} \int_{k'}^k dq \, \beta_q} G(k) \lesssim \frac{1}{(N \Delta \beta)^2} \int_0^{2\pi} dk G(k), \nonumber \\
    & G(k) = [(\mathcal{A}_k^1)^2 + (\mathcal{A}_k^2)^2] \mathcal{A}_k^3 + (\partial_k \mathcal{A}_k^1) \mathcal{A}_k^2 - \mathcal{A}_k^1 (\partial_k \mathcal{A}_k^2),
\end{align}
where the integral over gauge-potential components yields a nonextensive and $\beta$-independent factor. This is the result quoted in Eq.~\eqref{eq:genericScalingCorrections}.
\end{widetext}

\section{Dynamical adiabaticity}
\label{sec:dynamicalAdiabaticity}

\begin{figure}
    \includegraphics[width=\columnwidth]{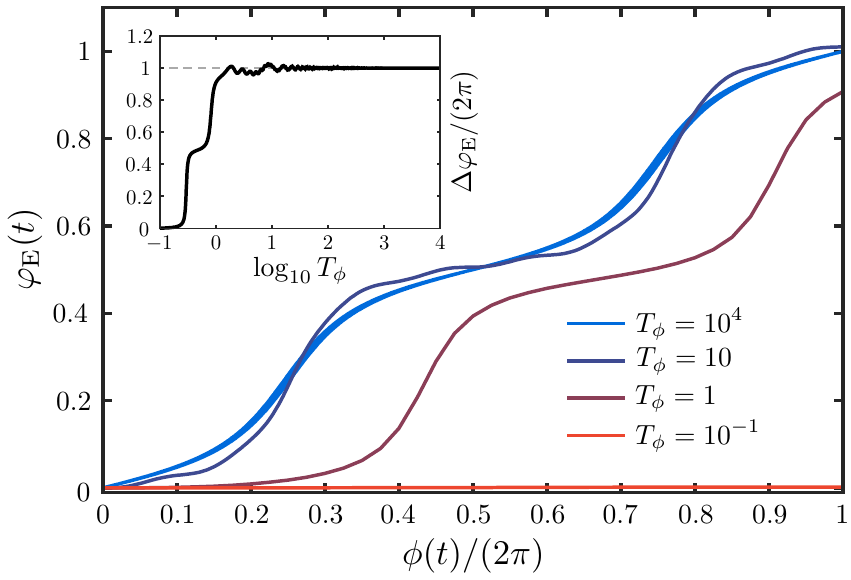}
    \caption{Dynamical measurement of the EGP $\varphi_\text{E}$ as a function of the parameter $\phi(t)/(2\pi) = t/T_\phi$ completing a closed cycle in parameter space over the time period $T_\phi$. The system examined here is the Rice-Mele model in thermal equilibrium discussed in Sec.~\ref{sec:examples}. \textbf{Inset}: Total EGP difference $\frac{1}{2\pi} \Delta \varphi_\text{E}$ accumulated over one cycle, as a function of $T_\phi$.}
    \label{fig:dynamicalAdiabaticity}
\end{figure}

If one chooses to measure the topological invariant $\frac{1}{2\pi} \Delta \varphi_\text{E} = 1/(2\pi) \oint d\phi \, \partial_\phi \varphi_\text{E}$ by observing the EGP $\varphi_\text{E}(\phi)$ \emph{continuously in time} while varying some system parameter $\phi$ along a closed loop, one must do so slowly enough to ensure that the state of the system follows the expected stationary state.

If one considers a \emph{nonequilibrium driven-dissipative} system with dissipative evolution governed by a Liouvillian, the rate at which the latter reaches its stationary state after an arbitrary change of parameters is controlled by the damping gap $\Delta_\text{d}$ (see Sec.~\ref{sec:examples}). In that case, the rate at which the EGP changes as $\phi$ is varied should be less than $\Delta_\text{d}$, i.e., $\vert \partial_t \varphi_\text{E} \vert \ll \Delta_\text{d}$. This translates as the ``dynamical adiabaticity'' condition
\begin{equation}
    \vert \dot{\phi} \vert \ll \vert \partial_\phi \varphi_\text{E} \vert^{-1}  \, \Delta_\text{d},
    \label{eq:dampingAdiabaticity}
\end{equation}
which should hold all along the path $\phi \in [0, 2\pi]$. Note that the ``rate'' of change $\vert \partial_\phi \varphi_\text{E} \vert$ generically increases when moving closer to purity-gap-closing points (recall that $\phi$ typically encircles such points). Therefore, the requirements for dynamical adiabaticity are determined by both the damping and the purity gaps.

If one considers, instead, a system at \emph{thermal equilibrium} with unitary evolution governed by a Hamiltonian, the above discussion cannot be applied directly. Unless one explicitly takes into account the coupling between the system and the reservoir(s) which make(s) the latter thermalize, the damping gap is not defined and, due to the finite temperature, there is no many-body energy gap either --- even if the chemical potential lies in the gap between two bands. As we have shown in the main text, however, the temperature-induced population of single-particle energy states above the chemical potential does not affect the EGP winding $\frac{1}{2\pi} \Delta \varphi_\text{E}$, at any finite temperature. Therefore, in the thermal case, we expect dynamical adiabaticity to be controlled by the energy gap $\Delta \epsilon$ between single-particle energy bands below and above the chemical potential, rather than by the damping gap as in Eq.~\eqref{eq:dampingAdiabaticity}. The fact that this is indeed correct is exemplified in Fig.~\ref{fig:dynamicalAdiabaticity} for the Rice-Mele model (see Sec.~\ref{sec:examples}) at half filling with $8$ sites and thermal initial state (temperature $T = 10$). The hopping amplitudes $t_1, t_2$ and the staggered potential $\Delta$ [see Eq.~\eqref{eq:RiceMeleModel}] are varied continuously in time so as to encircle the purity-gap-closing point $t_2 = t_1$, $\Delta = 0$. Namely, we parameterize $t_{1,2} = 2 \pm \sin(2\pi \phi/T_\phi)/4$ and $\Delta = -\cos(2\pi \phi/T_\phi)/2$, and vary $\phi$ linearly in time from $0$ to $2\pi$ over the time period $T_\phi$, i.e., $\phi = 2\pi t/T_\phi$. Figure~\ref{fig:dynamicalAdiabaticity} shows the EGP as function of $\phi(t)/(2\pi)$ for different values of $T_\phi$. As expected, the EGP difference $\frac{1}{2\pi} \Delta \varphi_\text{E}$ accumulated over one cycle is approximately quantized provided that $T_\phi$ is large as compared to the inverse energy gap $\Delta\epsilon = \mathcal{O}(1)$ (inset of Fig.~\ref{fig:dynamicalAdiabaticity}).

\bibliographystyle{apsrev4-1}
\bibliography{bibliography}

\end{document}